\documentclass[a4paper,11pt]{article}
\usepackage{jcappub} 

\arxivnumber{2302.07333} 
\title{Early dark energy constraints with late-time expansion marginalization}

\author[a]{João Rebouças,}
\author[b]{Jonathan Gordon,}
\author[a]{Diogo H. F. de Souza,}
\author[b]{Kunhao Zhong,}
\author[b]{Vivian Miranda,}
\author[a]{Rogerio Rosenfeld,}
\author[c]{Tim Eifler}
\author[c,d]{and Elisabeth Krause}
\affiliation[a]{Instituto de Física Teórica da Universidade Estadual Paulista and ICTP South American Institute for Fundamental Research, \\ R. Dr. Bento Teobaldo Ferraz, 271, Bloco II, Barra-Funda - São Paulo/SP, Brasil}
\affiliation[b]{C. N. Yang Institute for Theoretical Physics, Stony Brook University,\\ Stony Brook, NY 11794, USA}
\affiliation[c]{Department of Astronomy and Steward Observatory, University of Arizona,\\ 933 N Cherry Ave, Tucson, AZ 85719, USA}
\affiliation[d]{Department of Physics, University of Arizona,\\  1118 E Fourth Str, Tucson, AZ, 85721-0065, USA}

\emailAdd{joao.reboucas@unesp.br}

\abstract{
    Early dark energy (EDE) is an extension to the $\Lambda$CDM model that includes an additional energy density contribution near recombination. The model was proposed to reduce the tension between the measurements of the Hubble constant $H_0$ from the cosmic microwave background (CMB) and from the local cosmic distance ladder.
    Some analyses in the recent literature have shown intriguing hints for EDE.
    However, this model increases the tension in the derived clustering of galaxies (as measured by the so-called $S_8$ parameter) between CMB and large scale structure (LSS) measurements. 
    This new tension limits the contribution of EDE during recombination, and thus its effect on the Hubble tension.
    In this work, we investigate whether the inclusion of a general, smooth late-time dark energy modification can increase back the EDE contribution when LSS data is included in the analysis.
    In order to generalize the late expansion with respect to the $\Lambda$CDM model, we substitute the cosmological constant by a late dark energy fluid model with a 
    piecewise constant equation of state $w(z)$ in redshift bins. 
    We show that, when analysing this generalized model with combinations of CMB, LSS and type Ia supernovae data from several experiments no significant changes on $S_8$ and EDE parameter constraints is found. The contribution to
    the EDE fraction constraint with late-time expansion marginalization is $f_\mathrm{EDE} = 0.067^{+0.019}_{-0.027}$ using 3 redshift bins, with similar results for 5 and 10 redshift bins. 
    This work shows that in order to solve simultaneously the Hubble and $S_8$ tensions, one needs a mechanism for increasing the clustering of matter at late times different from a simple change in the background evolution of late dark energy.
}


\begin{document}
\maketitle
\flushbottom

\section{\label{sec:introduction}Introduction}

A viable cosmological model must be able to consistently explain the myriad of observational data stemming from a large span of the history of the Universe, from the formation of light nuclei in its first minutes to the distribution of galaxies formed in the past couple billion years. The spatially flat $\Lambda$CDM model has emerged as the standard cosmological model over the past decades: it is the simplest and most viable cosmological model (see e.g. \cite{Frieman:2008sn}). It is based on a cosmological constant $\Lambda$ to account for the current acceleration of the Universe \cite{riess-acceleration}, cold dark matter (CDM) necessary for structure formation and the existence of an early phase of inflation responsible for generating primordial perturbations with a power spectrum characterized by its amplitude and slope.

New physics beyond $\Lambda$CDM would manifest itself through tensions in the derived parameters from different observations as a consequence of the model incompleteness.
Potentially, we are starting to see the first hints of such incompleteness from the recent tensions in measuring the Hubble constant $H_0$.
Although this tension appears in several measurements of early-time and late-time observables (for a review, see \cite{tensions_early_late, cosmology_intertwined_h0}), it is most acute in the comparison between an indirect determination from the Cosmic Microwave Background (CMB) obtained by the Planck collaboration \cite{planck2018results} and local measurements using the cosmic distance ladder \cite{sh0es2018,sh0es21,h0_trgb_calibrated}, where a tension of the order of 5$\sigma$ is reported.

In addition to the discrepancy in $H_0$, a less severe tension has also been reported in a quantity related to the 
amplitude of matter fluctuations, the so-called $S_8$ parameter, with  $S_8 = \sigma_8\sqrt{\Omega_m/0.3}$, where $\sigma_8$ is the  RMS amplitude of matter fluctuations at the scale $8h^{-1}$Mpc and  $\Omega_m$ is the current matter density fraction \citep{cosmology_intertwined_s8}.
Measurements of two-point correlation functions in galaxy surveys such as the Dark Energy Survey \citep{desy3, desy3_shear_calibration, desy3_shear_modelling, desy3redmagic, desy3maglim, desy3-harmonic, desy1, desy1-shear}, the Kilo-Degree Survey \citep{kids, kids-450, kids1000, kids-1000-2} and the Hyper Suprime-Cam Subaru Strategic Program \citep{hsc-erratum, hsc-y1-harmonic, hsc-real-space}, report a significantly smaller value for $S_8$ than the Planck 2018 result. For instance, this difference is at a $2.3\sigma$ level with the cosmic shear analysis from DES in harmonic space \cite{desy3-harmonic}, while the recent analysis from KiDS reports a $3\sigma$ difference \citep{kids-1000-2}.

While these differences could be explained by unaccounted-for systematic errors in any of the experiments (see e.g. \cite{sne-color-calibration-hubble, nonlinear-solution-s8-tension, consistent_lensing_low_s8}), cosmological tensions could also be an indication that the standard $\Lambda$CDM model is unable to explain all data simultaneously, calling for new physics effects beyond this model. There has been a large body of research exploring different possibilities of extending the $\Lambda$CDM model in order to alleviate these tensions (see e.g. \cite{hubble_hunters_guide, h0_olympics, in-the-realm-of-the-hubble-tension, snowmass21-hubble-tension}).

Among these possibilities, the presence of an early dark energy (EDE) component,
proposed before the rise of the Hubble tension \citep{Wetterich:2004pv,doran-ede},
has been actively studied in this context (see e.g. \cite{de_early_times, fluidEDE, rocknroll, ede_can_restore}).
A non-negligible amount of dark energy around the recombination era that quickly dissipates afterwards would decrease the sound horizon relevant for CMB while leaving the diameter distance to the last scattering surface unchanged and 
hence could in principle ameliorate the Hubble tension \citep{edeshiftshorizon}.

EDE has been the subject of discussion in the recent literature, with ups and downs \cite{ups_and_downs_ede}. For instance, in \cite{ede_can_restore} a $2\sigma$ detection of EDE was reported in combined Planck 2018, SH0ES and Pantheon data, reducing the Hubble tension to $1.7\sigma$. Recent CMB data from the Atacama Cosmology Telescope (ACT) favors significantly the presence of EDE. ACT reports a $3\sigma$ preference for EDE in comparison with $\Lambda$CDM \cite{act_ede}, whereas
\citep{ede-spt3g} reports a $2.6\sigma$ evidence when combining ACT and SPT-3G and \citep{hints_ede_act_planck_spt} reports a $3.3\sigma$ evidence for EDE when combining ACT, SPT-3G and Planck 2018 data, 
even without the SH0ES late-time priors on $H_0$ \citep{ede-act-no-late-priors}. 
However, recent studies have shown that large-scale structure (LSS) data disfavors EDE because the model causes an increase in the $S_8$ posterior mean, aggravating the $S_8$ tension \citep{ede_not_restore, ede_with_lss, consistency-lcdm-early-sachs}. In such works, LSS data limits the amount of EDE during recombination to values insufficient to solve the Hubble tension.
While this is a strong argument against the EDE model, the discussion is not yet settled  \citep{ede-not-excluded-by-lss, ede-planck-eftboss, herold_ede, ede-microphysics}.

Moreover, modifications in the late-time expansion history can affect the growth of structure, such as dynamical, smooth dark energy models \cite{growth_rate_dark_energy, falsifying_smooth, testable_de_predictions, testing-de-paradigms-weak-lensing, vivian_cora_pca, ede-pre-and-post-recombination}. Several models combining EDE with other modifications were studied in order to tackle both $H_0$ and $S_8$ tensions, such as decaying dark matter \cite{ede_ddm}, interacting dark energy \cite{interactingde, chameleon-ede}, models with spatial curvature \cite{ede-and-curvature} and constant $w$ \cite{ede_wconst_lens}, models with a $(w_0,w_a)$ parameterization for the equation of state \cite{ede_w0wa_ruinzinho}, models with massive neutrinos \cite{ede_massive_nu} and models with massive neutrinos coupled to dark energy \cite{ede_coupled_massivenu}, models with an anti-de Sitter vacuum phase \cite{ede-and-ads} and axionic dark matter \cite{ede-ads-axion}, to mention a few. For a recent review see \cite{kamionkowski-ede}.

The main goal of this work is to test new EDE models with modifications in the late-time expansion history of the Universe using recent LSS data and how they affect $S_8$ and the EDE constraints. 
We will model late dark energy as a fluid with negligible perturbations inside the horizon, also known as smooth dark energy. 
Also, we choose to employ a general parametrization for the late-time equation-of-state $w(z)$ as a piecewise constant function over several redshift bins. 
The redshift bins are chosen for $z < 3$, when dark energy starts becoming relevant to the background expansion. This model goes beyond the usual $(w_0,w_a)$ parametrization, allowing for a more general study
that is more suited for our goal.
By marginalizing EDE constraints over the parameters describing $w(z)$, we obtain predictions that are independent of late-time expansion behaviour. 
We use different combinations of available cosmological datasets to assess the EDE parameter posteriors in these modified models. 
We primarily use CMB data from ACT which favors EDE models, and therefore allow us to better quantify the shifts in the 
posterior distributions of the EDE parameters. In addition, we use the recent DES-Y3 data on LSS in our analysis.

This paper is organized as follows: section~\ref{sec:models} describes the dark energy models used in our analysis; section~\ref{sec:datasets} describes the data used in our work and analysis methods; section~\ref{sec:results} display the results of the analysis; we present our conclusions in section~\ref{sec:conclusions}.

\section{Dark Energy Models}
\label{sec:models}

In this section we describe the models we adopt to describe early and late dark energy.

\subsection{Early Dark Energy (EDE)}
\label{sec:ede}
There are two approaches to model EDE: using a microscopic scalar field description or using an effective fluid description.
In the first case, EDE can be realized by a minimally-coupled scalar field and is fully described by its potential and initial conditions. It is usually assumed a generalized ``axion-like" potential of the form
\begin{equation}
    V(\phi) = m^2 f^2 \left[ 1 - \cos\left( \frac{\phi}{f}\right) \right]^n.
\end{equation}
\label{eq:axionpotential}
The usual axion potential is obtained for $n=1$, where in this case $m$ is the axion mass and $f$ is the scale of spontaneous symmetry breaking that gives rise to the axion as a pseudo-Nambu-Golsdtone boson. 
Initially, the field is frozen at its initial value by the large Hubble friction, with an equation of state $w_\phi \approx -1$. As the Universe expands, the Hubble friction decreases and at a critical redshift denoted by $z_c$ the field starts to oscillate around the minimum of the potential, with an average equation of state $w_n = (n-1)/(n+1)$ \citep{Turner:1983he}. For values of $n \geq 2$, the energy density in this field decays faster than matter and quickly becomes negligible with respect to the matter density. Although lacking a well-motivated fundamental origin, there are recent attempts to derive these models (for $n>1$) from an ultraviolet theory \citep{McDonough:2022pku}. The axion-like potential is used by most EDE analyses in the literature (e.g. \cite{ede_not_restore, act_ede, de_early_times, ede_with_lss, ede-act-no-late-priors}). 
A similar scalar field model is the so-called ``Rock 'n' Roll" EDE \cite{rocknroll}, where a power-law potential is adopted:
\begin{equation}
    V(\phi) = V_0 \left(\frac{\phi}{M_\mathrm{pl}}\right)^{2n},
\end{equation}
where $M_\mathrm{pl}$ is the reduced Planck mass. We have also analysed this class of scalar field EDE models in this work.
In order to do so we used the implementation of the scalar field perturbations in CAMB with different potentials.

Although exact, the scalar field description may suffer from numerical issues arising from the rapid oscillations around the minimum of the potential.
The corresponding equation of state also experience fast variation and in this case one may use an effective fluid description with time-averaged quantities. 
In the case of axion-inspired models of EDE an effective fluid description was developed in \cite{fluidEDE} and it is characterized by the critical redshift $z_c$ 
mentioned above and the energy density fraction of EDE at $z_c$, $f_{\mathrm{EDE}}(z_c) = \rho_{\mathrm{EDE}}(z_c)/\rho_{\mathrm{tot}}(z_c)$.
We parameterize the background quantities as:
\begin{equation}
   \rho_{\mathrm{EDE}}(z) = \frac{2\rho_{\mathrm{EDE}}(z_c)}{1 + \left( \frac{1+z_c}{1+z} \right)^{3(1+w_n)}},
\end{equation}
where $z_c$ and $w_n$ are defined as above.
The corresponding EDE equation of state $w_{\mathrm{EDE}}$ is given by:
\begin{equation}
    w_{\mathrm{EDE}}(z) = \frac{1 + w_n}{1 + \left( \frac{1+z}{1+z_c} \right)^{3(1+w_n)}} - 1.
\end{equation}
At early times ($z \gg z_c$), the EDE energy density is approximately constant with $w_{\mathrm{EDE}} \rightarrow  -1$ since the field is frozen and behaves as a cosmological constant, while for $z \ll z_c$, the energy density scales as $\rho_\mathrm{EDE} \propto (1+z)^{3(1+w_n)}$. We will assume $n=3$ in this work, since this is the preferred value determined in previous analyses \cite{ede_can_restore}, \cite{edesptpol}.

In order to study perturbations in the EDE fluid one also needs to specify the effective sound speed $c_s$ (the adiabatic sound speed $c_a$ is fixed by the background quantities). For a slow-rolling canonically normalized scalar field  $c_a^2=w$ and $c_s^2=1$ \citep{gdm-hu}. 
However, this is not valid anymore when the scalar field is rapidly oscillating. In this case, the adiabatic sound speed takes into account variations of the equation of state:
\begin{equation}
    c_a^2 = w_{\mathrm{EDE}} - \frac{\dot{w}_{\mathrm{EDE}}}{3(1+ w_{\mathrm{EDE}}) {\cal H}},
\end{equation}
where the overdot denotes derivative with respect to conformal time, and the effective sound speed is given by  (see \cite{fluidEDE}):
\begin{equation}
        c_s^2(k, a) = \frac{2a^2(n - 1)\bar{\omega}_0^2 a^{-6w_n} + k^2}{2a^2(n + 1)\bar{\omega}_0^2 a^{-6w_n} + k^2},
\end{equation}
which follows from averaging both background and perturbations of the scalar field over its fast oscillations around the minimum of the potential.

The function $\bar{\omega}_0$ depends on an parameter $\Theta_i$, which is related to the scalar field initial value by $\Theta_i = \phi_i/f$. Although this is a fluid model, it reproduces the oscillation-averaged behavior of the axion-like scalar field and therefore both models have the same number of parameters \cite{fluidEDE}.

In summary, the effective fluid description of EDE is characterized by three parameters: $z_c$, $f_{\mathrm{EDE}}(z_c)$ (which, for brevity, we will refer to as $f_{\mathrm{EDE}}$) and $\Theta_i$, which fixes the effective sound speed $c_s$.
In the following we will mostly use the effective fluid description of EDE and we will briefly compare our results with a field theoretical description of EDE using both the axion-inspired and the "Rock 'n' Roll" models.

The background and linear perturbation equations for the EDE models have been incorporated in the Einstein-Boltzmann code \textsc{camb}\footnote{\protect\url{github.com/cmbant/CAMB}}. For the scalar field evolution, we set the initial value of the field, $\phi_i$ at $a = 10^{-7}$, when the field is still frozen. When comparing results from the scalar field with the effective fluid description of EDE we use a shooting method implemented in \textsc{camb} to transform the potential parameters $m$ and $f$ (or $V_0$ and $\phi_i$ in the Rock 'n' Roll model) into the effective fluid parameters $z_c$ and $f_{\mathrm{EDE}}$.

\subsection{Late Dark Energy}
\label{sec:binw}
Late dark energy has a significant impact on the growth of perturbations and hence it affects the 
cosmological parameter $S_8$. In this work we want to assess if a general model for late dark energy can 
ameliorate the $S_8$ tension found in EDE models.
Models with a canonically normalized, slow-rolling scalar field have an effective sound speed $c_s=1$ and hence behave as smooth dark energy since in this case perturbations are suppressed inside the horizon.

Smooth dark energy models only affect the growth of structure through its contribution to the background expansion, determined by its equation of state $w(z)$. 
In order to be agnostic about the behavior of dark energy at late times, we consider a piecewise equation of state for different redshift bins:
\begin{equation}
    w(z) = \begin{cases}
        w_0 & \text{ if } 0 \leq z < z_1,\\
        ... \\
        w_{n-1} & \text{ if } z_{n-1} \leq z < z_n,\\
        w_n = -1 & \text{ if } z \geq z_n.
    \end{cases}
\end{equation}
\label{eq:binw}
This parameterization can approximate any dynamical dark energy model that does not vary rapidly within the redshift bins.

In this work we test three different bin choices such that have a significant number of redshift bins inside the Pantheon supernovae range of $z \in [0.01, 2.3]$, when late dark energy becomes relevant to the background energy density. The choices are:
\begin{itemize}
    \item $z_i = \{0.7, 1.4, 2.1\}$,
    \item $z_i = \{0.5, 1.0, 1.5, 2.0, 3.0\}$,
    \item $z_i = \{0.3, 0.6, 0.9, 1.2, 1.5, 1.8, 2.1, 2.4, 2.7, 3.0\}$.
\end{itemize}

We show that these three cases are sufficient to test the impact of a general late dark energy model on $S_8$. 
Adding extra parameters slows down the convergence of Monte Carlo Markov Chains (MCMCs), and thus we limit our investigation to 10 bins.

We implement both early and late dark energy models in a modified version of 
\textsc{camb}, which allows the combination of multiple dark energy fluids. 
The late dark energy perturbations are calculated using the parametrized post-Friedmann approach \citep{hu_ppf}. In this approach there are no instabilities in the perturbations from either phantom crossing ($w=-1$ crossing) or from discontinuities in the $w(z)$ which arise in our parametrization.

\section{Datasets and Analysis}
\label{sec:datasets}
In this work we assess posteriors for EDE parameters marginalized over
the behaviour of late dark energy using different datasets from CMB, SNe Ia, LSS and strong lensing measurements.
We detail below these datasets and present our analysis strategy.

\subsection{CMB data}

We choose to use ACT data as our primary CMB dataset because the strongest claim of EDE detection comes from its analysis\footnote{While this paper was in the referral process, a preprint appeared in the arXives reporting improved constraints on a similar axion-like early dark energy model using a new temperature and polarization Planck likelihood, finding $f_\mathrm{EDE} < 0.061$ \cite{ede-improved-constraint-planck}, without including DES data. We expect our results that smooth dark energy cannot be used to alleviate the $S_8$ tension to be valid in such a scenario as well. We left a detailed comparison between Planck and ACT data on EDE models fitted simultaneously with generalized late-time physics for future work. We thank the referee for bringing this reference to our attention.} \citep{act_ede}.
With ACT data, we can more clearly quantify the effects of late-time DE marginalization on the EDE constraints, the main focus of this work.
We use the ACT CMB TT, TE and EE power spectra included in the \textsc{actpol DR4} likelihood \cite{act_cosmo} and implemented in \textsc{pyactlike}\footnote{\protect\url{https://github.com/ACTCollaboration/pyactlike}}. The spectra are measured in the multipole range $\ell \in [600, 4500]$ for TT and $\ell \in [350, 4500]$ for TE and EE.  We restrict ourselves to the multipole range $\ell \leq 3000$ in order to avoid biases due to baryonic feedback processes, as explained in \cite{baryon_feedback}. The only nuisance parameter in the likelihood is the polarization efficiency $y_p$.

We complement ACT data with a subset of the Planck 2018 CMB data \cite{planck2018results}. Following \cite{act_ede}, we combine the ACT power spectra with the Planck 2018 TT power spectrum for $\ell < 650$, where Planck has more precise data compared to ACT.
ACT data by itself cannot constrain the reionization optical depth $\tau$. Instead of imposing a Gaussian prior on $\tau$ as in the ACT official analyses \cite{act_cosmo, act_ede}, we choose to include Planck 2018 large-scale polarization spectrum ($\ell < 30$) data in order to constrain $\tau$. The combination of ACT spectra up to $\ell = 3000$, Planck 2018 TT spectrum cut at $\ell = 650$ and low-$\ell$ EE will be henceforth referred to simply as CMB. We also study the inclusion of CMB Lensing from Planck 2018, and we refer to this dataset as CMBL.

\subsection{SNe Ia data}
We use the Pantheon sample of 1048 SNe Ia in the redshift range $0.01 < z < 2.3$ \citep{pantheon} to test 
the recent expansion history in our models\footnote{We use Pantheon rather than the very recent dataset Pantheon+ \cite{pantheonplus}. Pantheon is a subset of Pantheon+ with 1048 (of the 1550 in Pantheon+) spectroscopically confirmed Type-IA supernovae.
The cosmological constraints from both datasets are compatible within $1\sigma$. 
In addition, the SNIa data does not constrain $S_8$, one of the major purposes in our work. 
Thus, we do not expect significant changes in our results by using Pantheon+.
}. We refer to this dataset as SNe.

\subsection{Large-Scale Structure data}
We use Baryonic Acoustic Oscillation (BAO) data from multiple sources: the Six-degree Field Galaxy Survey (6dFGS) \cite{bao_sixdf}, the SDSS DR7 main galaxy sample \cite{sdss-dr7} and from SDSS BOSS DR12 consensus sample \cite{sdss-boss-dr12}. We refer to this combination simply as BAO. The galaxies included in these surveys range from $z = 0.1$ to $z = 0.6$. Thus, the evolution of the BAO feature can provide information about the low redshift expansion history. 

In addition, we use Dark Energy Survey (DES) measurements of two-point correlation functions of galaxy shear, galaxy clustering and their cross-correlations (the so-called 3x2pt analysis).
DES data provides detailed information about the clustering of matter at low redshifts, 
strongly constraining $S_8$ and $\Omega_m$ and complementing other geometrical probes of the Universe.
Since the most stringent constrains on $S_8$ comes from cosmic shear we also perform analyses with cosmic shear data only.
All data was obtained from the DES-Y1\footnote{\protect\url{https://des.ncsa.illinois.edu/releases/y1a1/key-products}} and  DES-Y3\footnote{\protect\url{https://des.ncsa.illinois.edu/releases/y3a2}} data release repositories.

We use the results from two galaxy catalogs: the source \textsc{metacalibration} shape catalog and the \textsc{redmagic} lens galaxy sample. The source shape catalog \cite{desy3sourcecatalog} contains the ellipticities and photometric redshifts of over 100 million galaxies and it is divided in 4 redshift bins, each bin with limits at $z = [0, 0.36, 0.63, 0.87, 2.0]$. The \textsc{redmagic} lens sample \citep{desy3redmagic} contains 2.5 million galaxies whose positions are used to measure clustering correlations and galaxy-galaxy lensing. This lens sample is divided in 5 redshift bins with limits at $z = [0.15, 0.35, 0.50, 0.65, 0.80, 0.90]$.

The galaxy clustering two-point angular correlation function in real space $w(\theta)$ is calculated in the following way \cite{desy3cov}: the lens sample of galaxies is divided in redshift bins, each bin $i$ with a galaxy number density $n_{\mathrm{l}}^i(z)$. The source galaxy sample is also divided in redshift bins, with density $n_\mathrm{s}^j(z)$. Let
\begin{equation}
    q_{\text{l}}^i(\chi) = b^i_\mathrm{L} \times  n^i_{\mathrm{l}}(z(\chi))  \frac{dz}{d\chi}
\end{equation}
be the radial weight function, where $b^i_\mathrm{L}$ the linear galaxy bias, assumed constant on each redshift bin and scale independent, and $\chi$ is the comoving radial distance. We can calculate the two-point angular correlation function in real space by summing the correlations in Fourier space:
\begin{equation}
    w^i(\theta) = \sum_\ell \frac{2\ell + 1}{4\pi}P_\ell(\cos\theta)C^{ii}_{\delta_g\delta_g}(\ell),
    \label{eq:wtheta}
\end{equation}
where $P_\ell(x)$ is the $\ell$-th Legendre polynomial. 

The Fourier space correlations are calculated using the Limber approximation:
\begin{equation}
    C^{ij}_{\delta_g \delta_g}(\ell) = \int \frac{d\chi}{\chi^2} q_\mathrm{l}^i\left( \chi\right) q_\mathrm{l}^j\left(\chi \right) P_{\mathrm{NL}}\left(k = \frac{\ell + 1\big/2}{\chi}, z(\chi)\right),
\end{equation}
where $P_{\mathrm{NL}}(k,z)$ is the nonlinear matter power spectrum. The calculation of the shear auto-correlations is performed in a similar way. Defining the lensing efficiency as:
\begin{equation}
    q_\kappa^i(\chi) = \frac{3\Omega_mH_0^2}{2} \times \int_\chi^{\chi_H} d\chi' \left(\frac{\chi' - \chi}{\chi}\right) n^i_\mathrm{s}(z(\chi')) \frac{dz}{d\chi'},
\end{equation}
we calculate the correlations in real space as a sum in Fourier space:
\begin{equation}
    \xi^{ij}_{\pm}(\theta) = \sum_\ell \frac{2\ell + 1}{4\pi} \frac{2(G^+_{\ell,2}(x) \pm G^-_{\ell,2}(x))}{\ell^2(\ell+1)^2}C^{ij}_{\kappa\kappa}(\ell),
\end{equation}
where $x = \cos\theta$ and $G^{\pm}$ are analytic functions described in appendix A from \cite{desy3cov}. The Fourier space correlations are given by:
\begin{equation}
    C^{ij}_{\kappa\kappa}(\ell) = \int \frac{d\chi}{\chi^2} q_\kappa^i(\chi) q_\kappa^j(\chi)P_{\mathrm{NL}}\left(k = \frac{\ell + 1\big/2}{\chi}, z(\chi)\right).
\end{equation}
The correlation between galaxy shear and lens position is given by:
\begin{equation}
    \gamma_t^{ij}(\theta) = \sum_\ell \frac{2\ell + 1}{4\pi}\frac{P_\ell^2(\cos\theta)}{\ell(\ell+1)}C^{ij}_{\delta_g\kappa}(\ell),
\end{equation}
where $P_\ell^2(x)$ is the associated Legendre polynomial $P_\ell^m(x)$ with $m=2$ and:
\begin{equation}
    C_{\delta_g\kappa}^{ij}(\ell) = 
    \int \frac{d\chi}{\chi^2} q_\mathrm{l}^i (\chi) q_\kappa^j(\chi) P_{\mathrm{NL}}
    \left(k = \frac{\ell + 1\big/2}{\chi}, z(\chi)\right).
\end{equation}

DES measures angular correlations averaged over a given angular bin. Therefore, one needs to modify the theory predictions to account for this procedure.
In order to average over a finite angular bin $[\theta_{\min}, \theta_{\max}]$, one assumes that galaxies are uniformly distributed. 
It can be shown that 
for the galaxy clustering correlation function $w(\theta)$ this leads to the modification in eq. (\ref{eq:wtheta}):
\begin{equation}
P_\ell(\cos \theta) \rightarrow \bar{P}_\ell(\cos \theta) = \frac{\left[P_{\ell+1}(x) - P_{\ell-1}(x) \right]_{\cos\theta_{\max}}^{\cos\theta_{\min}}}{(2\ell+1)(\cos\theta_{\min } - \cos\theta_{\max })},
\end{equation}
and the corresponding modifications for the galaxy-galaxy lensing correlation function $\gamma_t(\theta)$ and for the cosmic shear correlation functions $\xi_\pm(\theta)$ can be found in \cite{desy3cov}.

The angular separations $\theta$ in the correlation functions are divided into 20 logarithmically spaced bins between $2.5$ arcmin to $250$ arcmin. After scale cuts to mitigate unaccounted for nonlinear effects such as baryonic feedback from active galactic nuclei (AGN), the final data vector for the Y3 3x2 analysis has 533 elements.

DES analyses assume several modelling choices that are described and tested in \cite{modeling-strategy-validation-desy3}. 
A simple linear, scale-independent model is used to describe galaxy bias for each redshift bin, $\delta_g = b^i_\mathrm{L}\delta_m$. 
Intrinsic alignment is modelled using the Tidal Torquing and Tidal Alignment (TATT) model \citep{tidaltorquing}. 
Photometric redshift uncertainties in both the lens and source galaxy samples are modelled as off-sets in each bin $\Delta z^i_{\mathrm{l/s}}$. 
Uncertainties in the shear measurements are modelled as multiplicative bias parameters $m^i$. Yet another correction factor called point-mass parameter $B_i$
is introduced in the modelling of the galaxy-galaxy lensing correlation $\gamma_t$.
The DES nuisance parameters that are sampled and marginalized over in our analysis are summarized in Table~\ref{tab:priors_nuisance} \citep{desy3}.

We notice that the fiducial DES-Y3 3x2pt analysis uses an alternative lens galaxy sample, the so-called {\tt MagLim} sample \citep{desy3maglim}. 
The main reason is that the 2x2pt analysis (i.e. without the shear correlation functions) revealed 
 an unexplained internal tension which manifested itself in, among other places, the linear galaxy bias $b_L$ derived from galaxy clustering and galaxy-galaxy lensing \citep{desy3redmagic}. 
 This result was further scrutinized in the final DES-Y3 3x2pt analysis, where an extra parameter denoted by $X_{\mathrm{lens}}$
 was introduced to account for this tension in a phenomenological manner \citep{desy3}. The inferred cosmological parameters from the fiducial analysis in $\Lambda$CDM are consistent with the ones obtained from the \textsc{redmagic} lens sample with and without marginalizing over $X_{\mathrm{lens}}$.
We will corroborate these results in our case by showing consistency between a 3x2pt analysis with $X_{\mathrm{lens}}=1$ and a shear-only analysis, where the latter is insensitive to 
$X_{\mathrm{lens}}$.

\subsection{Strong Lensing}
The H0LiCOW collaboration \cite{h0licow-intro, h0licow-5, h0licow} measured time-delay distances from six quasar-lens systems, providing an $H_0$ constraint that agrees with local distance ladder measurements while being completely independent from those: $H_0 = 73.3^{+1.7}_{-1.8} \mathrm{km} \; \mathrm{s}^{-1} \; \mathrm{Mpc}^{-1}$. We refrain from using SH0ES priors in our chains, as their $H_0$ constraints are in severe tension with CMB constraints, which would lead to a degradation of the CMB fit and an artificial tightening of our constraints. 
Therefore we chose to use strong lensing measurements from H0LiCOW instead of local SH0ES measurements. 
We refer to using the H0LiCOW likelihood prior as SL.

\subsection{Analysis}
\label{sec:analysis}
We perform our analysis using the \textsc{cocoa}\footnote{\protect\url{https://github.com/CosmoLike/cocoa}} pipeline \cite{cocoa_desy1}, which combines the \textsc{cobaya}\footnote{\protect\url{https://cobaya.readthedocs.io/en/latest/}} sampler \cite{cobaya} with the \textsc{cosmolike}\footnote{\protect\url{https://github.com/CosmoLike/}} code \cite{cosmolike} for computing likelihoods for galaxy surveys using a halo-based theoretical covariance matrix \citep{cosmocov} that was validated in DES-Y1 \citep{DES2017tss} and DES-Y3 \citep{desy3cov} analyses.
We use the Limber approximation to calculate cosmic shear and galaxy-galaxy lensing correlations, but not for galaxy clustering. The likelihoods from the other independent datasets are combined assuming no correlations.

We use our implemention of both early and late dark energy models in a modified version of 
\textsc{camb} with a Halofit model for the nonlinear matter power spectrum \citep{Takahashi:2012em} supplemented by 
the Casarini prescription to account for the variable equation-of-state $w(z)$ model for late dark energy \cite{Casarini_2016, Casarini_2009}\footnote{Following \cite{act_ede}, we increase the standard \textsc{camb} accuracy in order to calculate higher order multipoles. We use the increased accuracy settings \texttt{AccuracyBoost} = 1.2 and \texttt{max\_l} = 10000.}.
Neutrinos are modelled using the default \textsc{camb} configuration, with an effective number of relativistic species at recombination $N_{\text{eff}} = 3.046$, and only one massive neutrino species with mass $m_\nu = 0.06$ eV. In order to validate our \textsc{camb} version, we have reproduced results from literature, specifically Fig. 2 from \cite{rocknroll}, Fig. 1 from \cite{ede_can_restore} and Fig. 14 from \cite{act_ede}.

We sample over the base cosmological parameters $\{ \theta_*, \Omega_bh^2, \Omega_ch^2, \ln(10^{10}A_s), n_s, \tau\}$ with large, uninformative priors. 
We also sample over the EDE parameters $\{ \log_{10}(z_c), f_\mathrm{EDE}, \Theta_i \}$\footnote{We follow \protect\cite{ede_can_restore} and use $\Theta_i = \phi_i/f$ as a parameter instead of the effective speed of sound of the fluid.} instead of sampling over the potential parameters $m$ and $f$. This choice is widely used in previous EDE analyses (for instance \cite{act_ede}) and we adopted it here since it allows for easy comparison with these works. The piecewise-constant equation of state in each redshift bin $\{ w_i \}$ are also sampled with wide priors. The priors for base cosmological parameters and dark energy parameters are listed in Table~\ref{tab:priors_cosmo}. We also sample over nuisance parameters from ACT, DES-Y1 and DES-Y3. The priors for nuisance parameters follow the respective collaboration papers \cite{act_cosmo, desy1, desy3}, and are shown in Table~\ref{tab:priors_nuisance}. Following the fiducial DES-Y3 analysis, we also analytically marginalize over a point-mass parameter describing small-scale effects in galaxy-galaxy lensing. We use the Gelman-Rubin convergence criterion $|R-1| < 0.05$ to stop the chains. We use \textsc{getdist}\footnote{\protect\url{https://getdist.readthedocs.io/en/latest/}} \cite{getdist} to analyse the chains, obtain marginalized statistics and plot confidence contours. We emphasize that the main goal of this work is not to obtain state-of-the-art EDE constraints or compare the EDE model against $\Lambda$CDM. Rather, our main goal is to assess if general late-time dark energy models can affect EDE constraints, shifting $f_\mathrm{EDE}$ and $S_8$ posterior distributions, especially after including DES-Y3 LSS data. Thus, we refrain from performing model comparison analyses such as Bayesian evidence and Akaike Information Criterion (AIC): we simply compare the posterior distributions for different models and assess the changes in mean and  variance.

In order to study the effects of different datasets on the estimation of cosmological parameters
we run MCMC chains with five dataset combinations (DCs):
\begin{itemize}
    \item DC1: CMB + BAO + SNe + CMBL;
    \item DC2: CMB + BAO + SNe;
    \item DC3: CMB + BAO + SNe + CMBL + SL;
    \item DC4: CMB + SNe + CMBL;
    \item DC5: CMB + SNe + SL.
\end{itemize}
We chose the combination DC1 as the baseline dataset. By comparing it with DC2 we assess the effect of including CMBL. In the same way, comparing DC1 with DC4 allows us to gauge the effect of including BAO in the analysis.
DC3 includes SL, which tends to worsen the CMB fit, since the two datasets are somewhat in tension. Finally, DC5 also combines the datasets that are in tension, CMB on one side and SNe and SL on the other side of $H_0$. In order to study the impact of DES data on the EDE constraints, we run chains with and without DES data, using cosmic shear only and full 3x2pt data, for each combination of the datasets above. We also compare results between Y1 and Y3 analysis, where we use the respective data vectors, covariances, scale cuts and prior choices.

For each dataset combination, we run chains with EDE+$\Lambda$ and the combined EDE+$w(z)$ model with 3, 5, and 10 bins. Including tests and validation, we have run roughly 160 chains for this project. We describe our results in the next section.
\renewcommand{\arraystretch}{1.08}
\begin{table}
	\centering
	\begin{tabular}{lcc} 
		\hline
		\textbf{Parameter} & \textbf{Prior}\\
		\hline
        \textbf{Base Cosmological Parameters}\\
        \hline
        $\Omega_c h^2$ & $\mathcal{U}$[0.001, 0.99]\\
        $\Omega_b h^2$ & $\mathcal{U}$[0.005, 0.1]\\
        $100\times\theta_*$ & $\mathcal{U}$[0.5, 10]\\
        $\ln(10^{10}A_s)$ & $\mathcal{U}$[1.61, 3.91]\\
        $n_s$ & $\mathcal{U}$[0.85, 1.1]\\
        $\tau$ & $\mathcal{U}$[0.01, 0.8]\\
        \hline
        \textbf{Dark Energy Parameters}\\
        \hline
		$\log_{10}(z_c)$ & $\mathcal{U}$[3, 4.3]\\
		$f_{\mathrm{EDE}}(z_c)$ & $\mathcal{U}$[0.001, 0.5]\\
		$\Theta_i$ & $\mathcal{U}$[0.1, 3.1]\\
        $w_0$ & $\mathcal{U}$[-3, -0.1]\\
        $w_{n\geq 1}$ & $\mathcal{U}$[-3, 1]\\
        \hline
	\end{tabular}
    \caption{Priors for cosmological and dark energy parameters sampled in our MCMCs. $\mathcal{U}[a,b]$ represents an uniform distribution with edges $[a,b]$.}
    \label{tab:priors_cosmo}
\end{table}

\begin{table}
	\centering
	\begin{tabular}{lcc} 
		\hline
		\textbf{Parameter} & \textbf{Prior}\\
		\hline
        \textbf{ACT Nuisance Parameter}\\
        \hline
        $y_p$ & $\mathcal{U}$[0.9,1.1]\\
	\hline
        \textbf{DES-Y3 Nuisance Parameters}\\
        \hline
        Photometric redshift offsets\\
        $\Delta z_{\mathrm{source}}^{1}$ & $\mathcal{N}$[0, 0.018]\\
        $\Delta z_{\mathrm{source}}^{2}$ & $\mathcal{N}$[0, 0.015]\\
        $\Delta z_{\mathrm{source}}^{3}$ & $\mathcal{N}$[0, 0.011]\\
        $\Delta z_{\mathrm{source}}^{4}$ & $\mathcal{N}$[0, 0.017]\\
        $\Delta z_{\mathrm{lens}}^{1}$ & $\mathcal{N}$[0.006, 0.004]\\
        $\Delta z_{\mathrm{lens}}^{2}$ & $\mathcal{N}$[0.001, 0.003]\\
        $\Delta z_{\mathrm{lens}}^{3}$ & $\mathcal{N}$[0.004, 0.003]\\
        $\Delta z_{\mathrm{lens}}^{4}$ & $\mathcal{N}$[-0.002, 0.005]\\
        $\Delta z_{\mathrm{lens}}^{5}$ & $\mathcal{N}$[-0.007, 0.010]\\
        Intrinsic Alignment (TATT)\\
        $a^{1,2}$ & $\mathcal{U}$[-5, 5]\\
        $\eta^{1,2}$ & $\mathcal{U}$[-5, 5]\\
        $b_{\mathrm{TA}}^1$ & $\mathcal{U}$[0, 2]\\
        Linear Galaxy Bias\\
        $b_1^i$ & $\mathcal{U}$[0.8, 3]\\
        Shear Calibration\\
        $m^1$ & $\mathcal{N}$[-0.006, 0.009]\\
        $m^2$ & $\mathcal{N}$[-0.020, 0.008]\\
        $m^3$ & $\mathcal{N}$[-0.024, 0.008]\\
        $m^4$ & $\mathcal{N}$[-0.037, 0.008]\\
        \hline
        \textbf{DES-Y1 Nuisance Parameters}\\
        \hline
        Photometric redshift offsets\\
        $\Delta z_{\mathrm{source}}^{1}$ & $\mathcal{N}$[-0.001, 0.016]\\
        $\Delta z_{\mathrm{source}}^{2}$ & $\mathcal{N}$[-0.019, 0.013]\\
        $\Delta z_{\mathrm{source}}^{3}$ & $\mathcal{N}$[0.009, 0.011]\\
        $\Delta z_{\mathrm{source}}^{4}$ & $\mathcal{N}$[-0.018, 0.022]\\
        $\Delta z_{\mathrm{lens}}^{1}$ & $\mathcal{N}$[0.008, 0.007]\\
        $\Delta z_{\mathrm{lens}}^{2}$ & $\mathcal{N}$[-0.005, 0.007]\\
        $\Delta z_{\mathrm{lens}}^{3}$ & $\mathcal{N}$[0.006, 0.006]\\
        $\Delta z_{\mathrm{lens}}^{4}$ & $\mathcal{N}$[0, 0.01]\\
        $\Delta z_{\mathrm{lens}}^{5}$ & $\mathcal{N}$[0, 0.01]\\
        Intrinsic Alignment (NLA)\\
        $a^{1}$ & $\mathcal{U}$[-5, 5]\\
        $\eta^{1}$ & $\mathcal{U}$[-5, 5]\\
        Linear Galaxy Bias\\
        $b_1^i$ & $\mathcal{U}$[0.8, 3]\\
        Shear Calibration\\
        $m^i, \; i \in [1,4]$ & $\mathcal{N}$[0.012, 0.023]\\
        \hline
	\end{tabular}
    \caption{Priors for nuisance parameters sampled in the MCMC. $\mathcal{U}[a,b]$ represents an uniform distribution over the real interval $[a,b]$, while $\mathcal{N}[a,b]$ represents a Gaussian distribution with mean $a$ and standard deviation $b$.}
    \label{tab:priors_nuisance}
\end{table}

\section{Results and Discussion}
\label{sec:results}
This section is organized as follows: in section~\ref{sec:results-ede}, we present results for EDE without late-time dark energy modifications; in section~\ref{sec:results-late-marg} we show the results marginalized over late-time expansion, done by including the piecewise $w(z)$ model. A Principal Component Analysis of the late-time dark energy $w(z)$ is presented in appendix~\ref{sec:pca}.

\subsection{Early Dark Energy}
\label{sec:results-ede}
We start by discussing constraints on early dark energy models with the usual cosmological constant as late dark energy.
The results in this subsection expand on previous analyses with the inclusion of DES-Y3 data and the mitigation of baryonic effects on ACT data by considering multipoles $\ell < 3000$ in the TT, TE and EE power spectra.  

One of the main arguments given by previous works in the literature in favor of the EDE model is its ability to improve the CMB fit with respect to the $\Lambda$CDM model. The EDE analysis from the ACT collaboration \cite{act_ede} has reported a 3$\sigma$ evidence for EDE using full ACT spectra, Planck TT spectrum cut at $\ell = 650$, CMB lensing and BAO, finding $f_\mathrm{EDE} = 0.091^{+0.020}_{-0.036}$. 
This is in agreement with our results using the DC1 dataset, which differs from \cite{act_ede} by the inclusion of SNe data and the cut in the multipoles in the ACT power spectra. 

\begin{figure}
    \centering
    \includegraphics[width=.75\columnwidth]{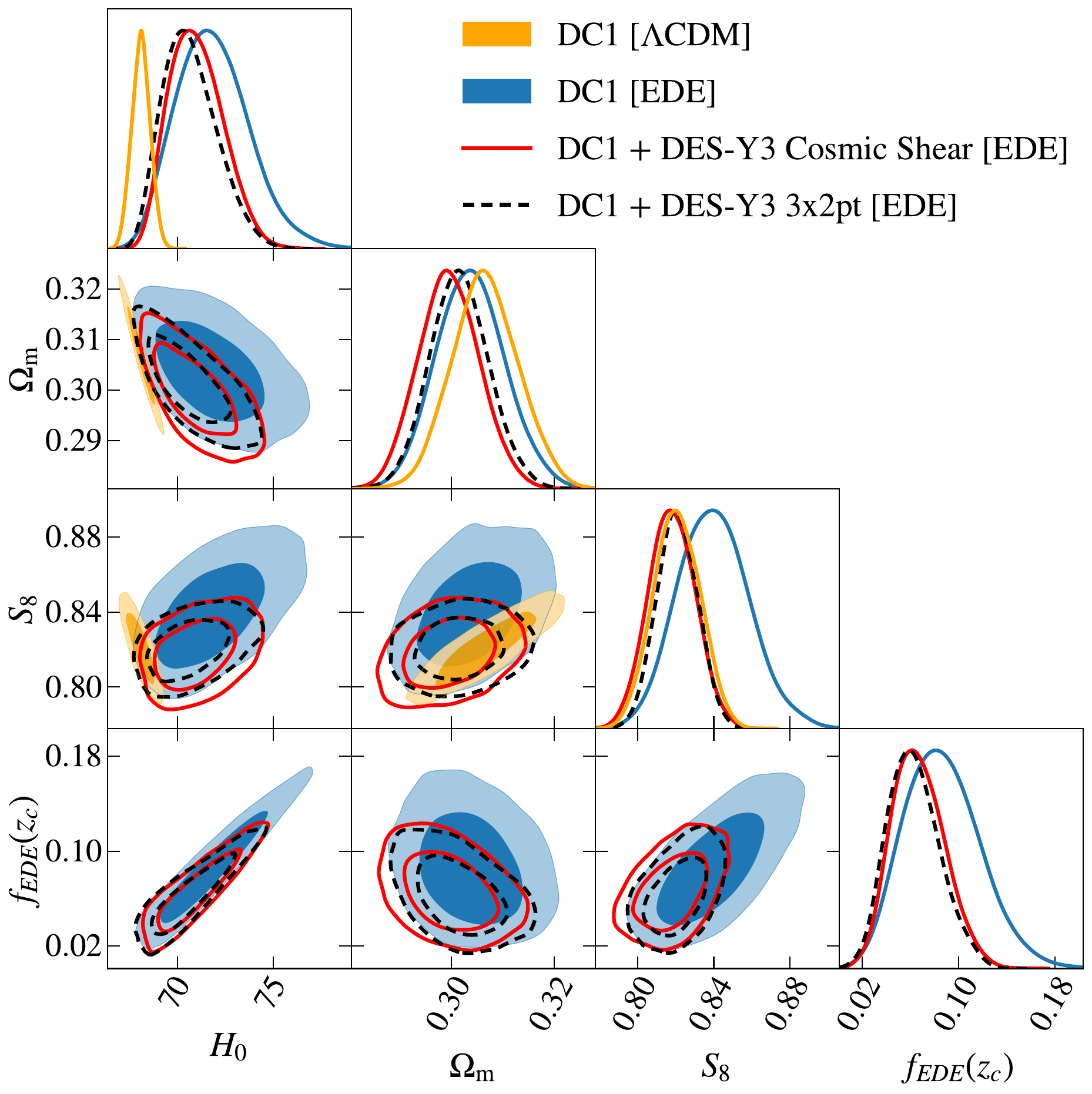}
    \caption{Constraints for the $\Lambda$CDM and effective fluid early dark energy models using the dataset combination DC1: ACT TT, TE, EE spectra with $\ell_\mathrm{max} = 650$, Planck 2018 TT spectrum with $\ell_\mathrm{max} = 650$ + Planck 2018 low-$\ell$ EE spectrum, BAO from 6dFGS, SDSS DR7 main galaxy sample and SDSS BOSS DR12 consensus sample, Pantheon supernovae and Planck 2018 CMB Lensing spectrum. We also include DES-Y3 3x2-point correlation function as well as cosmic shear alone.}
    \label{fig:ede_fluid_des_y3}
\end{figure}

Figure~\ref{fig:ede_fluid_des_y3} shows our results for the posterior distributions for the parameters $H_0$, $\Omega_m$, $S_8$ and $f_\mathrm{EDE}$, using both the EDE model and a standard $\Lambda$CDM model for the base dataset DC1; we find $f_\mathrm{EDE} = 0.088^{+0.028}_{-0.035}$.
The figure shows that a preference for nonzero EDE has the desirable property of relaxing the Hubble tension by increasing the $H_0$ posterior mean.
On the other hand, one notices that EDE also leads to a less dramatic increase in the $S_8$ posterior mean, from $S_8 = 0.820 \pm 0.012$ (DC1, $\Lambda$CDM) to $S_8 = 0.840^{+0.018}_{-0.020}$ (DC1, EDE). This effect further increases the disagreement with galaxy survey results, which prefer a significantly lower value in the $\Lambda$CDM scenario: the DES-Y3 fiducial analysis, for instance, reports $S_8 = 0.776 \pm 0.017$ \cite{desy3}.
In order to verify the effects of different implementations of EDE models, we show in Table \ref{tab:1dconstraints-edemodels}
the consistency among the marginalized constraints for the parameters $H_0$, $S_8$ and $f_\mathrm{EDE}$ obtained using dataset DC1
from the three different implementations of EDE models, namely an effective fluid description and the two scalar field models described in section~\ref{sec:ede}. We verified that this consistency remains when including DES data. In the following most of our analyses are performed in the effective fluid description.

\renewcommand{\arraystretch}{1.2}
\begin{table}
	\centering
	\begin{tabular}{lccc} 
		\hline
		\textbf{Model/Dataset} & $H_0 \; (\mathrm{km}\; \mathrm{s}^{-1}\; \mathrm{Mpc}^{-1})$ & $10 \times S_8$ & $100 \times f_\mathrm{EDE}$ \\
		\hline
        \textbf{DC1}\\
        \hline
        $\Lambda$CDM & $68.1\pm 0.50$ & $8.20\pm 0.12$ & --\\
        Axion-like & $71.6^{+1.3}_{-2.3}$ & $8.35^{+0.17}_{-0.22}$ & $10.6^{+2.7}_{-4.9}$\\
        Rock 'n' Roll & $71.4^{+1.3}_{-2.0}$ & $8.34^{+0.16}_{-0.19}$ & $9.5^{+2.7}_{-3.6}$\\
        Effective Fluid & $71.9^{+1.6}_{-2.1}$ & $8.41^{+0.17}_{-0.20}$ & $8.8^{+2.8}_{-3.5}$ \\
        \hline
	\end{tabular}
    \caption{Marginalized constraints (mean and 68\% limits) for the parameters $H_0$, $S_8$ and $f_\mathrm{EDE}$ for three EDE models: Axion-like EDE, Rock 'n' Roll EDE and effective fluid EDE. We use the dataset combination CMB (ACT power spectra with $\ell_\mathrm{max} = 3000$, Planck 2018 TT spectrum with $\ell_\mathrm{max} = 650$ and EE spectrum with $\ell_\mathrm{max} = 30$) + BAO (6dFGS, SDSS DR7 MGS and SDSS BOSS DR12 consensus sample) + SNe (Pantheon) + CMBLens (Planck 2018 CMB Lensing potential).}
    \label{tab:1dconstraints-edemodels}
\end{table}

We can assess the goodness-of-fit of each model by comparing the values of $\chi^2$ at the minimum. We find that the difference between $\Lambda$CDM and EDE scenarios is concentrated in the CMB part of the data vector: $\chi^{2\; (\Lambda\mathrm{CDM})}_{\mathrm{CMB}} = 966.3 $ and $ \chi^{2\;(\mathrm{EDE})}_{\mathrm{CMB}} = 957.0 $, with difference $\Delta\chi^2_{\mathrm{CMB}} = \chi^{2\;(\mathrm{EDE})}_{\mathrm{CMB}} - \chi^{2\; (\Lambda\mathrm{CDM})}_{\mathrm{CMB}} = -9.3$. We remark that our CMB data vector consists of 911 components. Assuming $\Delta\chi^2$ is $\chi^2$-distributed with 3 degrees of freedom due to the extra EDE parameters, this difference corresponds to approximately a $2.2\sigma$ preference for EDE over $\Lambda$CDM, or a $97.5\%$ significance.

The simplest way of quantifying tensions in the parameter constraints between two analyses is by the 1D standardized difference \cite{tensions_lensing_planck, tension_metrics_des}:
\begin{equation}\label{eq:tension}
   T = \frac{|\mu_1 - \mu_2|}{\sqrt{\sigma_1^2 + \sigma_2^2}},
\end{equation}
where $\mu_i$ is the marginalized mean value of the parameter posterior distribution for each analysis and $\sigma_i$ is the marginalized 68\% confidence limit. If the posteriors are skewed, we take the average of the upper and lower limits. We remark that this simple tension metric does not capture the multidimensional correlations between different parameters and it only reflects a statistical significance when the posteriors are approximately Gaussian, which is the case for $H_0$ and $S_8$ but not for $f_\mathrm{EDE}$. 

\begin{figure}
    \centering
    \includegraphics[width=\columnwidth]{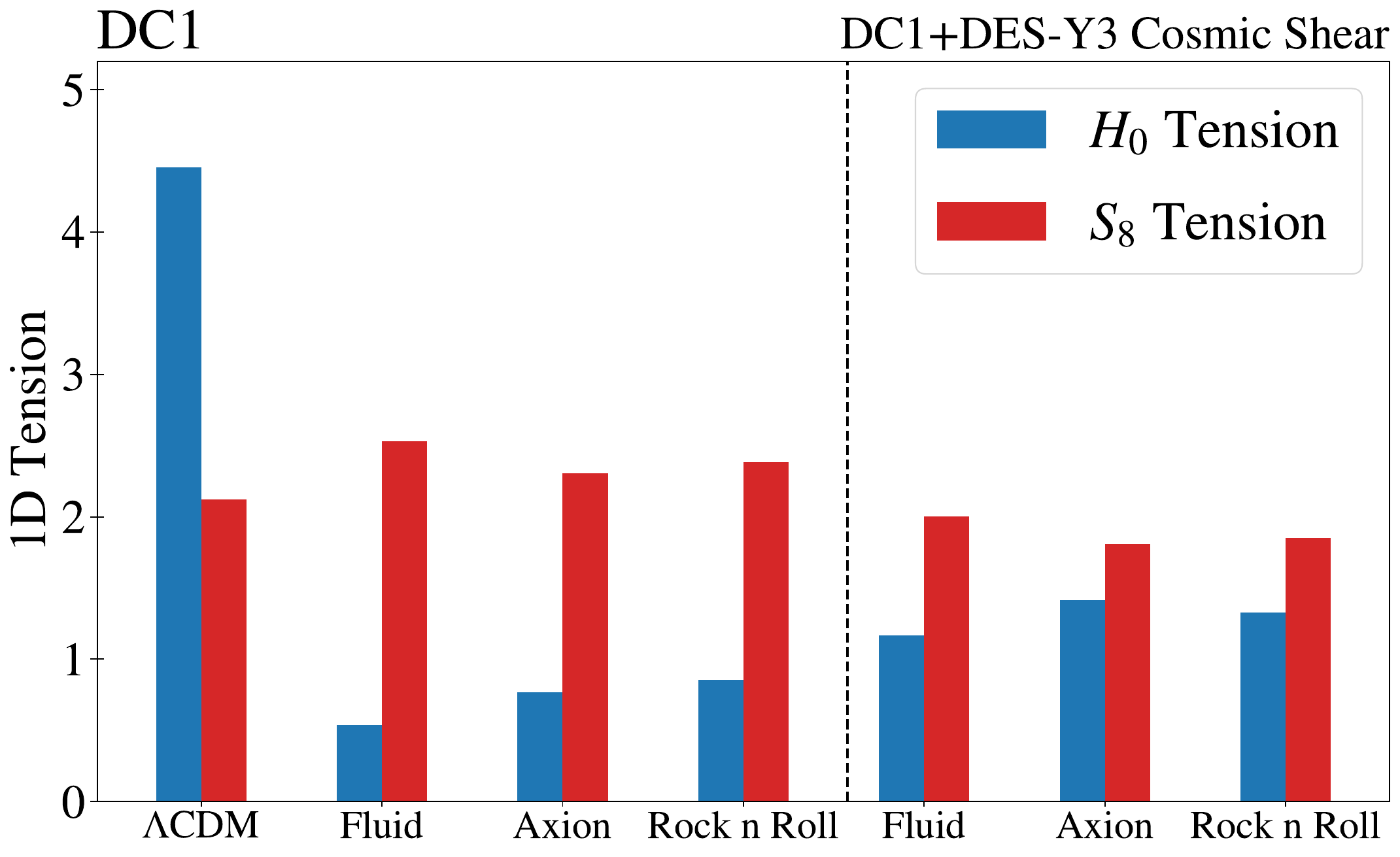}
    \caption{1D tensions between our analysis and \protect\cite{sh0es-gaia-august2022} (for the $H_0$ tension) or \protect\cite{desy3} (for the $S_8$ tension). To the left of the vertical dashed line, we are using the dataset combination DC1 ACT TT, TE, EE spectra with $\ell_\mathrm{max} = 650$, Planck 2018 TT spectrum with $\ell_\mathrm{max} = 650$ + Planck 2018 low-$\ell$ EE spectrum, BAO from 6dFGS, SDSS DR7 main galaxy sample and SDSS BOSS DR12 consensus sample, Pantheon supernovae and Planck 2018 CMB Lensing spectrum; to the right, we add DES-Y3 cosmic shear data. The horizontal axis represents different early dark energy model choices.}
    \label{fig:tensions}
\end{figure}

In Fig.~\ref{fig:tensions} we show the tensions arising from our analysis compared with a recent $H_0$ constraint reported by SH0ES \cite{sh0es-gaia-august2022} and the DES-Y3 constraints for $S_8$ \cite{desy3}, using the simple 1D metric of Equation~\ref{eq:tension}. As expected, 
the Hubble tension is decreased from roughly $4.5\sigma$ in $\Lambda$CDM to less than $1\sigma$ in EDE models and 
the different EDE models are in reasonable agreement.  
Regarding the $S_8$ parameter, one notices that the $S_8$ tension increases from roughly $2\sigma$ to $2.3-2.5\sigma$ when EDE is included. Thus, data from galaxy surveys should limit the amount of EDE in order to accommodate lower $S_8$ values.
We find similar results for the other datasets that we show as robustness tests in appendix~\ref{sec:dc345}.


We now turn our attention to the effects of including DES data on EDE constraints. Previous works such as \cite{ede_not_restore} and \cite{ede_with_lss} concluded that the evidence for EDE decreases when including galaxy clustering or cosmic shear data. Such works have used Planck 2018 CMB data for primary CMB and DES-Y1 or KiDS data for large-scale structure. We confirm their conclusions by reanalysing early dark energy models with recent data from DES-Y3.
A word of caution is needed since, as mentioned earlier, there is an internal tension in DES-Y3 analysis using the \textsc{redmagic} lens sample, which could introduce biases in the dark energy parameter constraints. For this reason in addition to the complete 3x2pt analysis we also present an analysis using cosmic shear data only, which is insensitive to this tension that appears only in the clustering amplitude.

Figure~\ref{fig:ede_fluid_des_y3} shows confidence contours using the dataset DC1 combined with DES-Y3 data. More details are presented in Table~\ref{tab:1d-constraints-dc1-dc2}, where we show the marginalized constraints on the parameters $H_0$, $S_8$ and $f_\mathrm{EDE}$ for DC1 and DC2 comparing the effect of including DES-Y3 3x2pt and cosmic shear-only data.
We observe no significant differences between 3x2pt and cosmic shear only constraints and thus conclude that the \textsc{redmagic} internal tension does not significantly affect the EDE constraints. We plan a future in-depth analysis comparing EDE constraints from each of the galaxy lens samples and using a $X_\mathrm{lens}$ bias correction factor \cite{desy3} in order to assess this robustness.

 Due to the $S_8$ tension and its positive correlation with $f_\mathrm{EDE}$, 
 adding large-scale structure data decreases the EDE contribution and worsens the Hubble tension. For the combination DC1, we find:
 \begin{align}
 \begin{split}
     f_\mathrm{EDE} &= 0.087^{+0.028}_{-0.035} \text{ (DC1)},\\
     f_\mathrm{EDE} &= 0.067^{+0.019}_{-0.025} \text{ (DC1+DES-Y3 } \xi_{\pm}),\\
     f_\mathrm{EDE} &= 0.063^{+0.019}_{-0.024} \text{ (DC1+DES-Y3 3x2pt).}
 \end{split}
 \end{align}
As expected, including DES-Y3 data results in a significant decrease on the estimated amount of EDE during recombination, disfavoring $f_\mathrm{EDE}$ values which solve the Hubble tension, i.e. around $10\%$.

We also compare our results of including either Y1 or Y3 data in our analyses below. 
For the combination DC1, the EDE constraints are
 \begin{align}
 \begin{split}
     f_\mathrm{EDE} &= 0.072^{+0.022}_{-0.029} \text{ (DC1+DES-Y1 } \xi_{\pm}),\\
     f_\mathrm{EDE} &= 0.057^{+0.018}_{-0.023} \text{ (DC1+DES-Y1 3x2pt).}
 \end{split}
 \end{align}
In the Y1 analysis, we observe a small difference in the posterior means between 3x2pt and cosmic shear constraints, as the former imposes tighter constraints on $f_\mathrm{EDE}$ than the latter. This difference is smaller in the Y3 analysis. 
The same happens for the combination DC2, without CMB lensing:
 \begin{align}
 \begin{split}
     f_\mathrm{EDE} &= 0.081^{+0.022}_{-0.031} \text{ (DC2+DES-Y1 } \xi_{\pm}),\\
     f_\mathrm{EDE} &= 0.064^{+0.017}_{-0.024} \text{ (DC2+DES-Y1 3x2pt),}\\
     f_\mathrm{EDE} &= 0.075^{+0.019}_{-0.030} \text{ (DC2+DES-Y3 } \xi_{\pm}),\\
     f_\mathrm{EDE} &= 0.068^{+0.020}_{-0.026} \text{ (DC2+DES-Y3 3x2pt).}
 \end{split}
 \end{align}

\begin{figure}
    \centering
    \includegraphics[width=.48\textwidth]{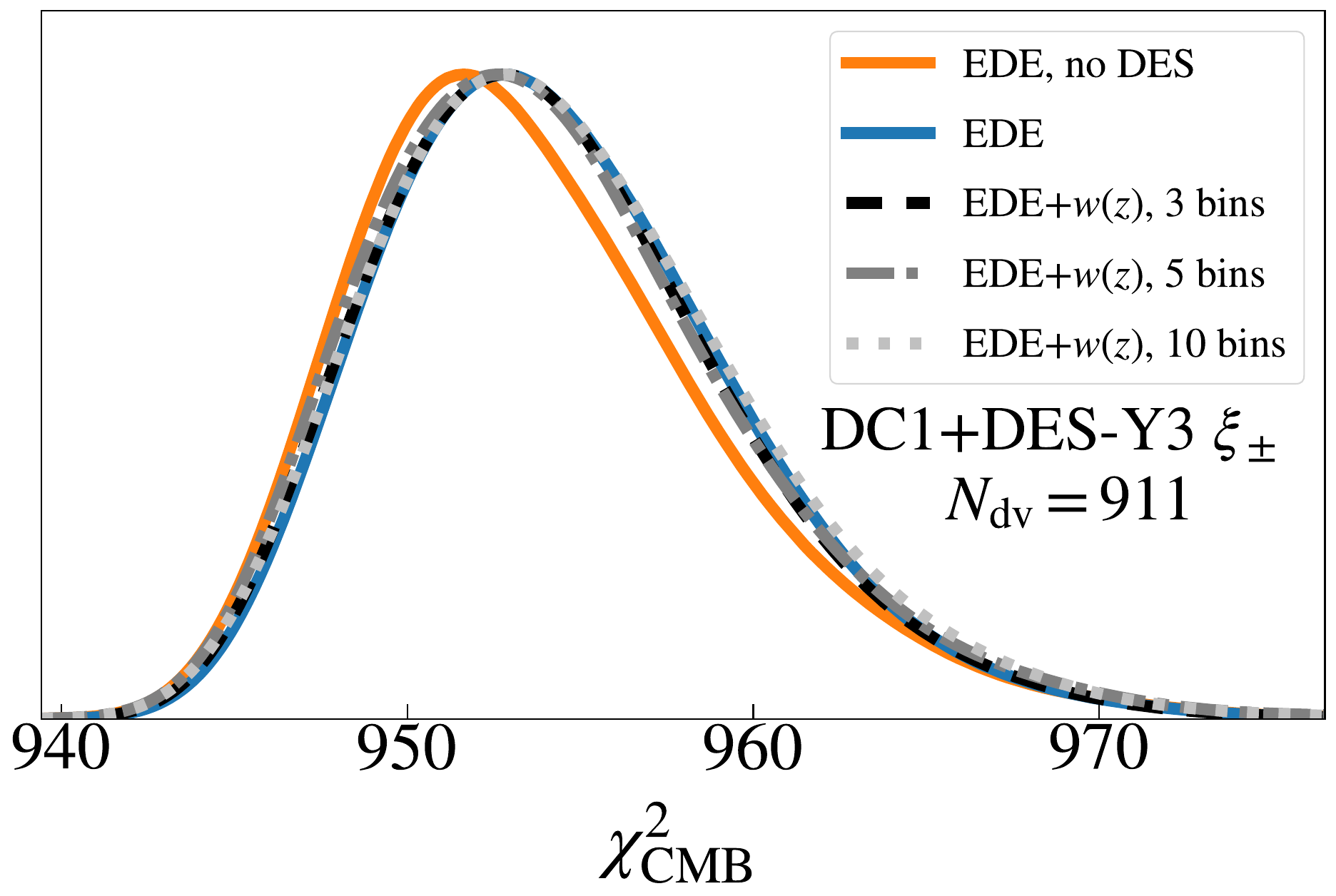}
    \includegraphics[width=.48\textwidth]{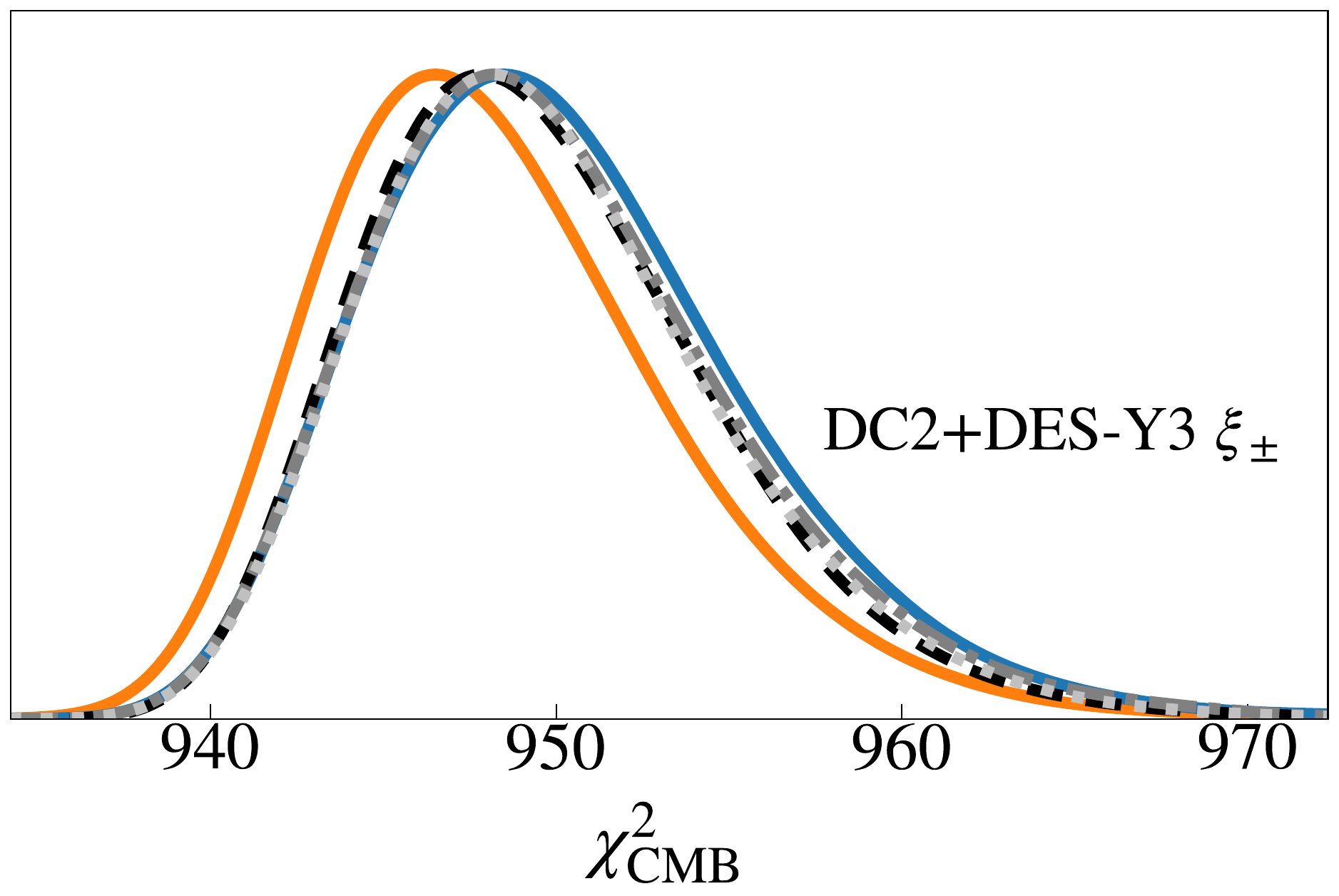}
    \caption{$\chi_\mathrm{CMB}^2$ (911 points) distributions in the MCMCs using the dataset combinations DC1 (ACT TT, TE, EE spectra with $\ell_\mathrm{max} = 650$, Planck 2018 TT spectrum with $\ell_\mathrm{max} = 650$ + Planck 2018 low-$\ell$ EE spectrum, BAO from 6dFGS, SDSS DR7 main galaxy sample and SDSS BOSS DR12 consensus sample, Pantheon supernovae and Planck 2018 CMB Lensing spectrum) and DC2 (same as DC1 without CMB Lensing), combined with DES-Y3 cosmic shear data, under the effective fluid EDE model and EDE + $w(z)$ with 3, 5 and 10 bins. We also include EDE results without DES data.}
    \label{fig:chi2-dc12}
\end{figure}

The inclusion of DES-Y3 3x2pt correlation functions shifts $H_0$ by $-0.6\sigma$ (increasing the tension with SH0ES data) and $S_8$ by $-1.3\sigma$ (decreasing the tension with DES-Y3-only constraints on $S_8$). Figure~\ref{fig:tensions} shows that the inclusion of DES-Y3 data worsens the $H_0$ tension solution, now above $1\sigma$ level.
Figure~\ref{fig:chi2-dc12} also shows that the inclusion of DES data slightly worsens the CMB fit. This effect is the main motivation to further extend the EDE scenario. In the following section, we explore whether substituting smooth dark energy for the cosmological constant can increase again the EDE contribution that results in a smaller $H_0$ tension. 

\subsection{Marginalization over Late-time Expansion}
\label{sec:results-late-marg}
In this section we are interested in the study of models with modifications to the standard $\Lambda$CDM model that account simultaneously for both early dark energy and a generic late dark energy parametrization, the latter characterized by a piecewise constant equation of state in different redshift bins. This parametrization will be tested with the different datasets described in section~\ref{sec:analysis}.

Analysing the dark energy equation of state in detail during the so-called rise of dark energy age \citep{rise-of-de} is an ambitious task by itself. Since this study is not directly related to EDE, we describe it in appendix~\ref{sec:pca}, where we show an estimation of the reconstruction of the late dark energy equation of state and a principal component analysis on its components $w_i$.

Since we are interested in the possible impact of these models on the EDE parameter estimations together with the $H_0$ and $S_8$ tensions, we focus our attention on the results obtained from marginalizing over the late-time dark energy parameters.
More specifically, we show results marginalized for 3, 5 and 10 redshift bins of constant $w(z)$, using 
the effective fluid EDE model with $n = 3$. 

\begin{figure*}
   \includegraphics[width=\textwidth]{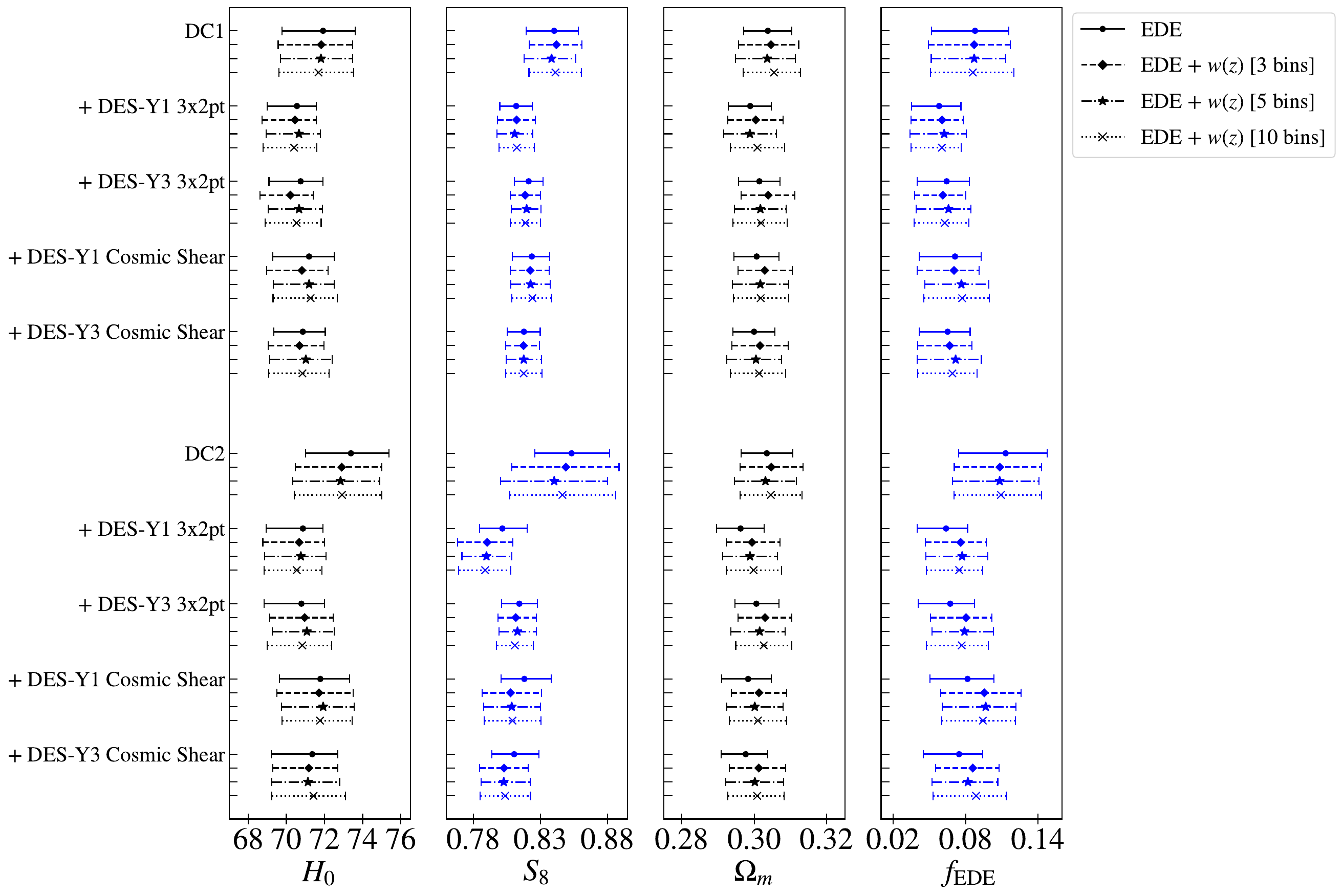}
    \caption{Marginalized constraints for $H_0$ (in $\mathrm{km} \; \mathrm{s}^{-1} \; \mathrm{Mpc}^{-1}$), $S_8$, $\Omega_m$ and $f_\mathrm{EDE}$, using the dataset combinations DC1 (ACT TT, TE, EE spectra with $\ell_\mathrm{max} = 650$, Planck 2018 TT spectrum with $\ell_\mathrm{max} = 650$ + Planck 2018 low-$\ell$ EE spectrum, BAO from 6dFGS, SDSS DR7 main galaxy sample and SDSS BOSS DR12 consensus sample, Pantheon supernovae and Planck 2018 CMB Lensing spectrum) and DC2 (same as DC1 without CMB Lensing). For each combination, we add DES 3x2-point correlation functions and cosmic shear-only data. DES-Y1 and DES-Y3 are shown in order to assess the impacts of the updated data set and analysis in our results. We use the effective fluid EDE model presented in section~\ref{sec:ede} combined with a late-time fluid with piecewise equation of state, described in section~\ref{sec:binw}, using 3, 5 and 10 redshift bins. We observe no significant shifts in the parameter constraints due to the modification of smooth dark energy at late times, regardless of the number of redshift bins in our model.}
    \label{fig:1dconstraints-priors12}
\end{figure*}

Our investigations are summarized in Fig.~\ref{fig:1dconstraints-priors12}, which shows marginalized 1D constraints (mean and 68\% limits) for the parameters $H_0$, $S_8$ and $f_\mathrm{EDE}$ for the dataset combinations DC1 and DC2, without and with DES data and for three different number of bins for late dark energy. Results for DES-Y1 and DES-Y3 are shown in order to assess the impacts of the updated data set and analysis in our results. The values for the posterior means and $68\%$ limits are shown in Table~\ref{tab:1d-constraints-dc1-dc2}. Constraints for the datasets DC3, DC4 and DC5 are shown in appendix~\ref{sec:dc345} as robustness tests. 

For DC1, DC3, DC4 and DC5 combined with DES data, we have found that adding the binned $w(z)$ model does not make a significant difference in $f_\mathrm{EDE}$ constraints, regardless of the number of redshift bins.

\renewcommand{\arraystretch}{1.2}
\begin{table}
	\centering
	\begin{tabular}{lccc} 
		\hline
		\textbf{Model/Dataset} & $H_0 \; (\mathrm{km}\; \mathrm{s}^{-1}\; \mathrm{Mpc}^{-1})$  & $10 \times S_8$ & $100 \times f_\mathrm{EDE}$ \\
		\hline
        \textbf{DC1}\\
        \hline
        $\Lambda$CDM & $68.09^{+0.49}_{-0.49}$ & $8.21\pm 0.13$ & --\\
        EDE & $71.9^{+1.6}_{-2.1}$ & $8.40^{+0.18}_{-0.20}$ & $8.7^{+2.8}_{-3.5}$\\
        EDE + 3-bin $w(z)$ & $71.8^{+1.7}_{-2.3}$ & $8.41\pm 0.19$ & $8.7^{+2.9}_{-3.8}$\\
        EDE + 5-bin $w(z)$ & $71.8^{+1.7}_{-2.1}$ & $8.38\pm 0.19$ & $8.7^{+2.8}_{-3.5}$ \\
        EDE + 10-bin $w(z)$ & $71.6^{+1.8}_{-2.1}$ & $8.41^{+0.18}_{-0.20}$ & $8.4^{+3.2}_{-3.6}$ \\
        \hline
        \small{\textbf{DC1+DES-Y3 $\xi_{\pm}$}}\\
        \hline
        EDE & $71.0^{+1.2}_{-1.6}$ & $8.18\pm 0.12$ & $6.7^{+1.9}_{-2.5}$\\
        EDE + 3-bin $w(z)$ & $70.6^{+1.3}_{-1.7}$ & $8.17 \pm 0.13$ & $6.7^{+1.9}_{-2.7}$\\
        EDE + 5-bin $w(z)$ & $71.0^{+1.3}_{-1.9}$ & $8.17\pm 0.13$ & $7.2^{+2.0}_{-3.1}$\\
        EDE + 10-bin $w(z)$ & $70.9^{+1.4}_{-1.8}$ & $8.18\pm 0.13$ & $7.0^{+2.1}_{-2.9}$\\
        \hline
        \textbf{DC1+DES-Y3 3x2pt}\\
        \hline
        EDE & $70.7^{+1.1}_{-1.6}$ & $8.21\pm 0.10$ & $6.3^{+1.9}_{-2.4}$\\
        EDE + 3-bin $w(z)$ & $70.3^{+1.3}_{-1.6}$ & $8.19\pm 0.11$ & $6.3^{+1.9}_{-2.5}$\\
        EDE + 5-bin $w(z)$ & $70.7^{+1.2}_{-1.6}$ & $8.19\pm 0.11$ & $6.6^{+1.9}_{-2.4}$\\
        EDE + 10-bin $w(z)$ & $70.6^{+1.2}_{-1.6}$ & $8.19\pm 0.11$ & $6.4^{+2.0}_{-2.6}$\\
        \hline
        \textbf{DC2}\\
        \hline
		$\Lambda$CDM & $68.06^{+0.56}_{-0.56}$ & $8.22\pm 0.17$ & --\\
        EDE & $73.3^{+2.1}_{-2.5}$ & $8.53^{+0.27}_{-0.31}$ & $11.2^{+3.6}_{-4.3}$\\
        EDE + 3-bin $w(z)$ & $73.0^{+2.1}_{-2.6}$ & $8.49\pm 0.41$ & $11.0^{+3.3}_{-4.1}$\\
        EDE + 5-bin $w(z)$ & $72.8^{+2.0}_{-2.5}$ & $8.38\pm 0.39$ & $10.7^{+3.3}_{-3.9}$ \\
        EDE + 10-bin $w(z)$ & $72.9^{+2.1}_{-2.5}$ & $8.46\pm 0.40$ & $10.9^{+3.4}_{-3.9}$ \\
        \hline
        \textbf{DC2+DES-Y3 $\xi_{\pm}$}\\
        \hline
        EDE & $71.4^{+1.3}_{-2.1}$ & $8.11 \pm 0.17$ & $7.5^{+1.9}_{-3.0}$\\
        EDE + 3-bin $w(z)$ & $71.2^{+1.6}_{-2.0}$ & $8.03^{+0.19}_{-0.17}$ & $8.6^{+2.4}_{-3.2}$\\
        EDE + 5-bin $w(z)$ & $71.2^{+1.6}_{-1.9}$ & $8.02^{+0.20}_{-0.17}$ & $8.4^{+2.4}_{-3.2}$\\
        EDE + 10-bin $w(z)$ & $71.3^{+1.6}_{-2.2}$ & $8.03\pm 0.19$ & $8.7^{+2.5}_{-3.5}$\\
        \hline
        \textbf{DC2+DES-Y3 3x2pt}\\
        \hline
        EDE & $70.9^{+1.3}_{-1.9}$ & $8.15\pm 0.13$ & $6.8^{+2.0}_{-2.6}$\\
        EDE + 3-bin $w(z)$ & $70.9^{+1.5}_{-1.8}$ & $8.12^{+0.15}_{-0.13}$ & $8.0^{+2.1}_{-2.9}$\\
        EDE + 5-bin $w(z)$ & $71.1^{+1.5}_{-1.8}$ & $8.13\pm 0.14$ & $7.9^{+2.6}_{-2.6}$\\
        EDE + 10-bin $w(z)$ & $71.0^{+1.5}_{-1.9}$ & $8.12 \pm 0.14$ & $7.9^{+2.2}_{-2.9}$\\
        \hline
	\end{tabular}
    \caption{Marginalized constraints (mean and 68\% limits) for the parameters $H_0$, $S_8$ and $f_\mathrm{EDE}$ for the dataset combinations DC1 (ACT TT, TE, EE spectra with $\ell_\mathrm{max} = 650$, Planck 2018 TT spectrum with $\ell_\mathrm{max} = 650$ + Planck 2018 low-$\ell$ EE spectrum, BAO from 6dFGS, SDSS DR7 main galaxy sample and SDSS BOSS DR12 consensus sample, Pantheon supernovae and Planck 2018 CMB Lensing spectrum) and DC2 (same as DC1 without CMB Lensing). We also combine both P1 and Base with DES-Y3 3x2-point correlations and cosmic shear.}
    
    \label{tab:1d-constraints-dc1-dc2}
\end{table}

Interestingly, for the DC2 dataset (i.e. without CMB lensing) combined with DES data, the inclusion of a dynamical smooth dark energy fluid at late-times does slightly affect the $S_8$ and $f_\mathrm{EDE}$ constraints. The effect, although very small, can be seen in all DES analyses and redshift binning choices. As shown in Table~\ref{tab:1d-constraints-dc1-dc2}, when including the binned $w(z)$ model, the $f_\mathrm{EDE}$ posterior mean is increased by $0.01$. This corresponds to a $0.5\sigma$ shift, well within the error bars. The $S_8$ posterior mean also slightly decreases by approximately $0.4\sigma$, which improves the fit to DES data. Figure~\ref{fig:chi2-dc12} also shows that, for the combination of DC2 and DES-Y3 cosmic 
shear, the late-time modification can also slightly improve the CMB fit. However, this increase in $f_\mathrm{EDE}$ does not come with an increase in $H_0$: we observe no significant shift in $H_0$ posterior. It seems the $w(z)$ models that are able to decrease $S_8$ also tend to decrease $H_0$. As CMB lensing further constrains the dark energy equation of state in redshifts $z < 3$, no $f_\mathrm{EDE}$ shifts can be detected in the other dataset combinations.

\section{Conclusions}
\label{sec:conclusions}
Cosmological tensions in $H_0$ and, to a lesser extent, in $S_8$, are challenging the standard $\Lambda$CDM model. If not explained by systematic effects, they may point to new physics. The presence of a non-negligible amount of dark energy around the recombination era, postulated in early dark energy models, has become a candidate solution for the Hubble tension, especially in light of new CMB data from ACT and SPT-3G. Cosmic shear and galaxy clustering data, however, disfavor the presence of EDE, since they aggravate the $S_8$ tension with CMB. This has been observed in previous analysis using Planck 2018, DES-Y1, KiDS and HSC data \cite{ede_not_restore, ede_with_lss}. Our work extends previous ones in several directions: incorporating the recent DES-Y3 data, analysing different combination of datasets and modifying the late dark energy model by allowing a smooth (unperturbed) dark energy completely characterized by its equation of state.

We also find that including recent large-scale strucuture data from DES-Y3 decreases the contribution for EDE and that this
is robust for different combinations of datasets. For
the combination DC1 we find $f_\mathrm{EDE} = 0.087^{+0.028}_{-0.035}$, compared to $f_\mathrm{EDE} = 0.067^{+0.019}_{-0.025}$ when DES-Y3 cosmic shear data is included, and $f_\mathrm{EDE} = 0.063^{+0.019}_{-0.023}$ using DES-Y3 3x2pt data. This effect makes EDE more unlikely as a Hubble tension solution. Our main conclusion is that smooth, late-time dark energy modifications cannot restore the EDE contribution to its significance without large-scale structure data. Using the DC1 combination and DES-Y3 cosmic shear data, we find $f_\mathrm{EDE} = 0.067^{+0.019}_{-0.027}$ for 3 redshift bins, $f_\mathrm{EDE} = 0.072^{+0.020}_{-0.031}$ for 5 bins and $f_\mathrm{EDE} = 0.070^{+0.021}_{-0.029}$ for 10 bins. Varying the number of redshift bins
does not add information to early-time dark energy constraints.

If we are to consider EDE as a solution to the Hubble tension, we must address the $S_8$ tension with either systematic effects or theoretical models beyond the smooth dark energy paradigm, for instance changing the mechanisms for clustering of matter, as done in \cite{ede_ddm, dark_sector_restore, step_in_s8}. In future works, we will analyse the combination between EDE and more general dark energy paradigms such as including anisotropic stress, modifying the sound speed, adding interactions between dark energy and CDM, among others.
Our results drives future research to other mechanisms affecting $S_8$ beyond the smooth dark energy paradigm combined with early dark energy models.

\appendix
\section{Constraints for Dataset Combinations 3, 4 and 5}
\label{sec:dc345}
\begin{figure*}
    \centering
    \includegraphics[width=\textwidth]{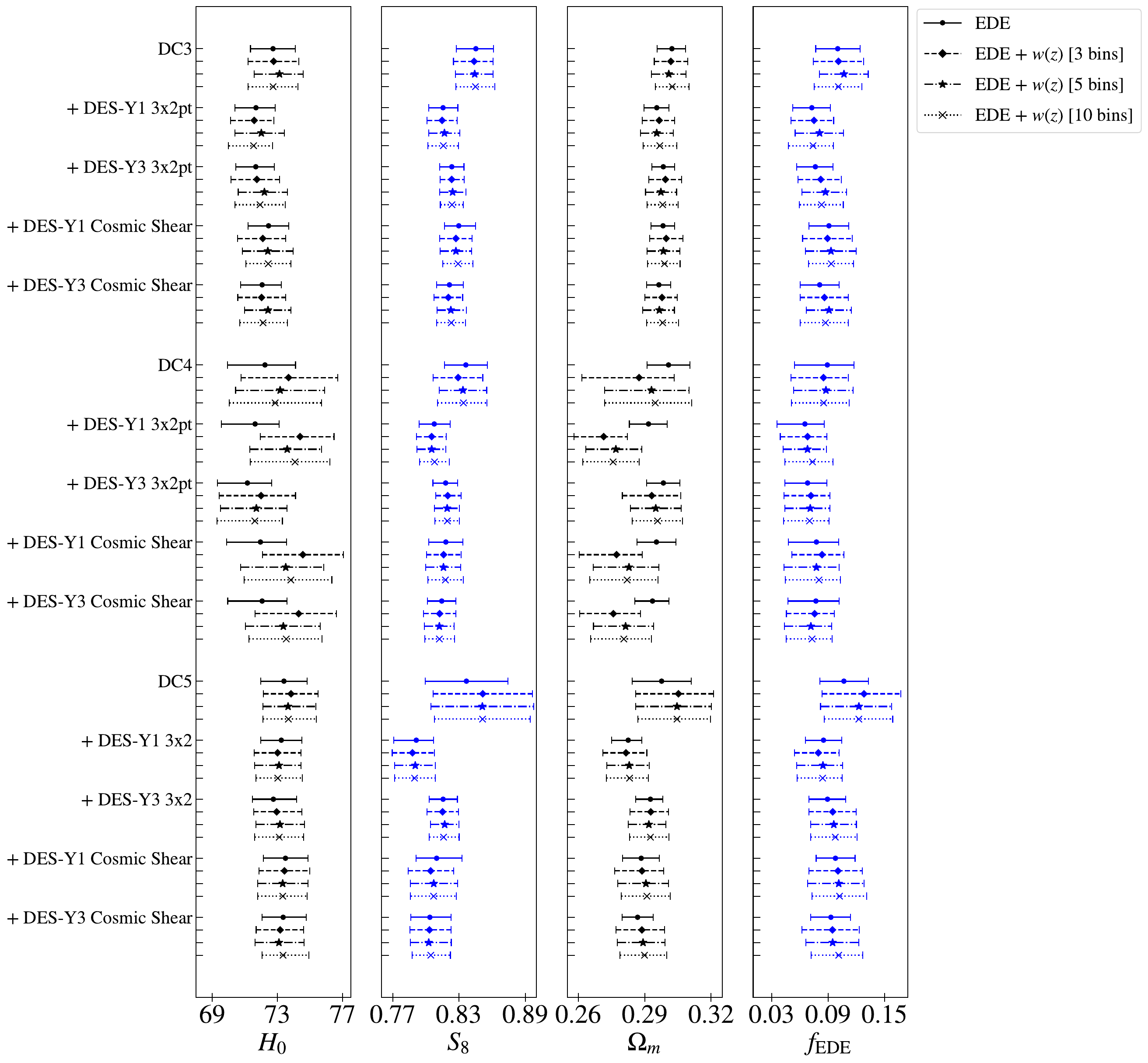}
    \caption{Marginalized constraints for $H_0$, $S_8$, $\Omega_m$ and $f_\mathrm{EDE}$, using the dataset combinations DC3 (ACT TT, TE, EE spectra with $\ell_\mathrm{max} = 650$, Planck 2018 TT spectrum with $\ell_\mathrm{max} = 650$ + Planck 2018 low-$\ell$ EE spectrum, BAO from 6dFGS, SDSS DR7 main galaxy sample and SDSS BOSS DR12 consensus sample, Pantheon supernovae, Planck 2018 CMB Lensing spectrum and H0liCOW strong lensing time delays), DC4 (same as DC3 without BAO and strong lensing) and DC5 (same as DC3, without BAO and CMB Lensing). For each combination, we add DES 3x2-point correlation functions and cosmic shear-only data. DES-Y1 and DES-Y3 are shown in order to assess the impacts of the updated data set and analysis in our results. We use the effective fluid EDE model presented in section~\ref{sec:ede} combined with a late-time fluid with piecewise equation of state, described in section~\ref{sec:binw}, using 3, 5 and 10 redshift bins. We observe no significant shifts in the $f_\mathrm{EDE}$ constraints due to the modification of smooth dark energy at late times, regardless of the number of redshift bins in our model.}
    \label{fig:1d-priors-345}
\end{figure*}

\begin{figure*}
    \centering
    \includegraphics[width=0.45\textwidth]{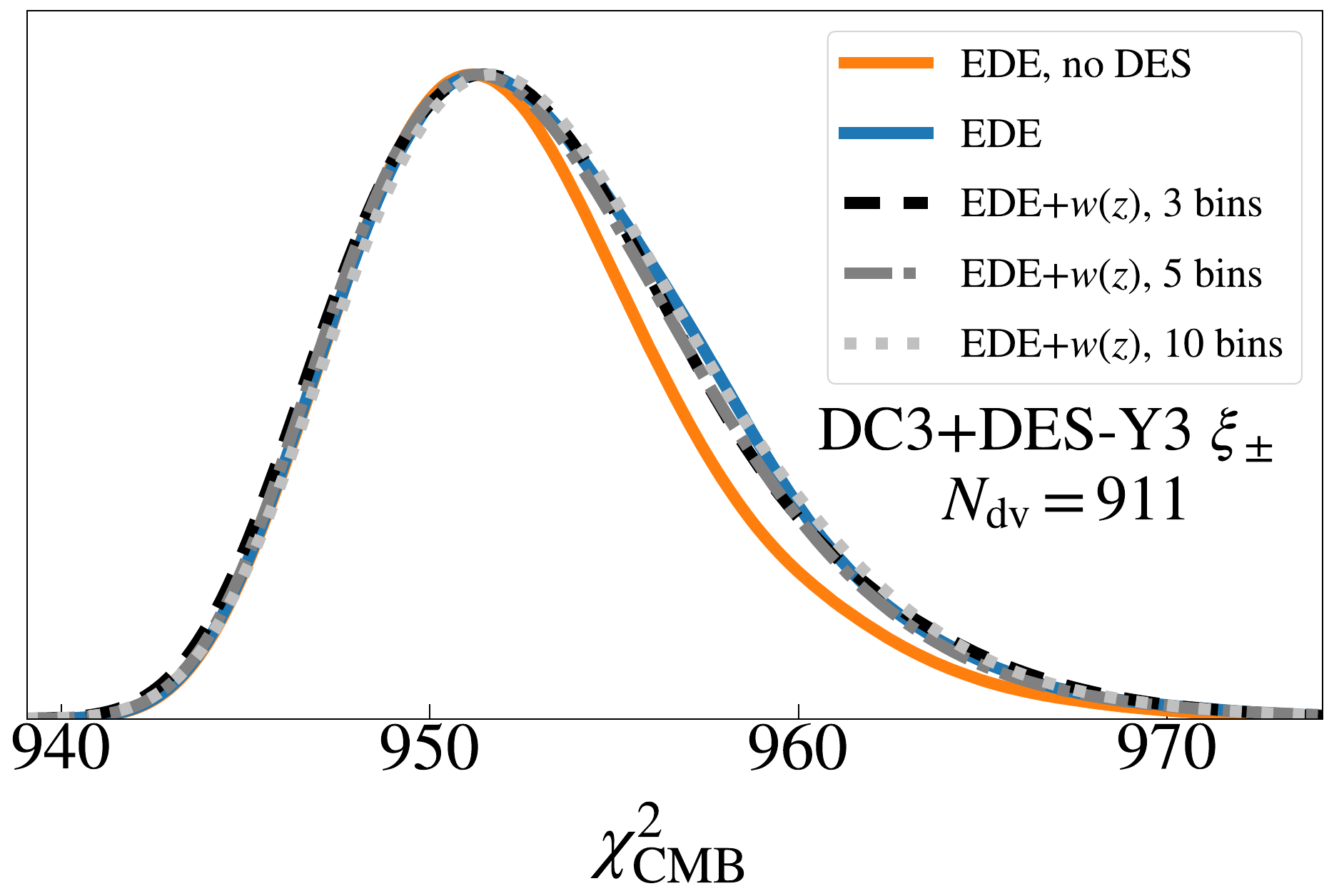}
    \includegraphics[width=0.45\textwidth]{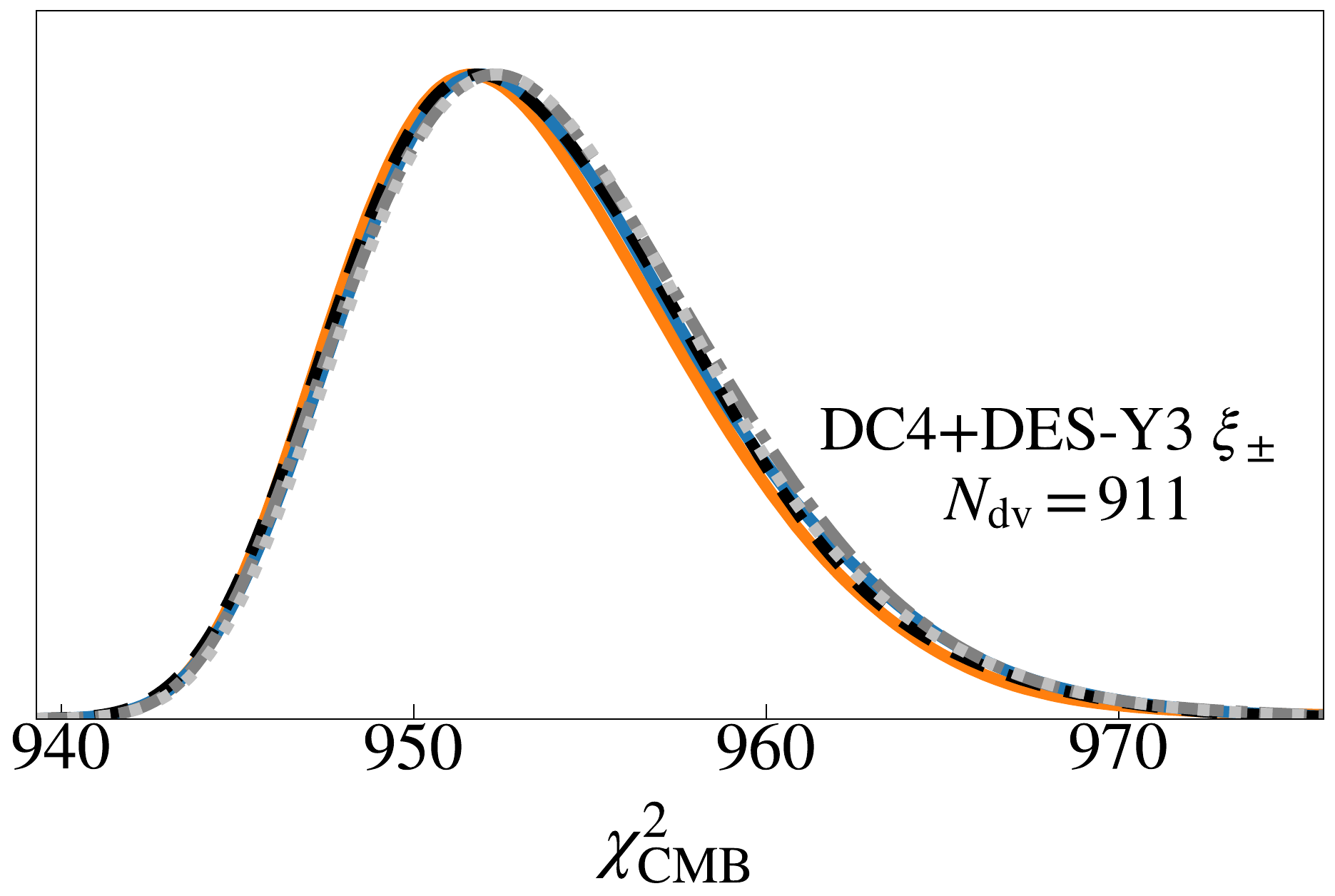}\\
    \includegraphics[width=0.45\textwidth]{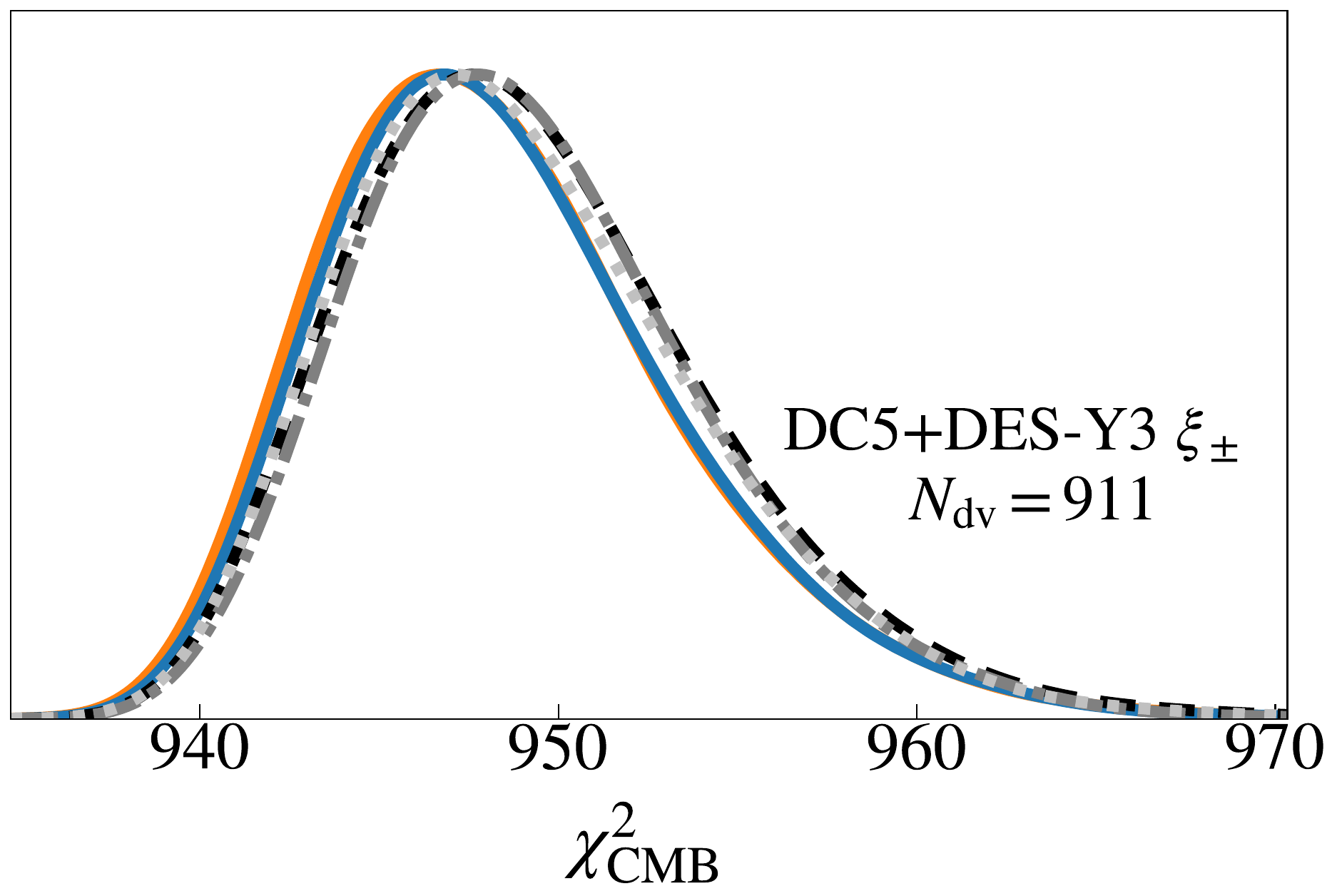}
    \caption{$\chi_\mathrm{CMB}^2$ (911 points) distributions in the MCMCs using the dataset combinations DC3 (ACT TT, TE, EE spectra with $\ell_\mathrm{max} = 650$, Planck 2018 TT spectrum with $\ell_\mathrm{max} = 650$ + Planck 2018 low-$\ell$ EE spectrum, BAO from 6dFGS, SDSS DR7 main galaxy sample and SDSS BOSS DR12 consensus sample, Pantheon supernovae, Planck 2018 CMB Lensing spectrum and H0liCOW strong lensing time delays), DC4 (same as DC3 without BAO and strong lensing) and DC5 (same as DC3, without BAO and CMB Lensing), combined with DES-Y3 cosmic shear data, under the effective fluid EDE model and EDE + $w(z)$ with 3, 5 and 10 bins. We also include EDE results without DES data.}
    \label{fig:chi2-binw-dc345}
\end{figure*}

In this appendix, we show results for DC3, DC4 and DC5, which corroborate our conclusions from the main text. In all dataset combinations, adding DES data does limit the amount of EDE during recombination and worsens the CMB fit. Furthermore, including the late-time modification does not restore the EDE detection to the significance it had without DES data. Figure~\ref{fig:1d-priors-345} shows the 1D marginalized constraints for the parameters $H_0$, $\Omega_m$, $S_8$ and $f_\mathrm{EDE}$, including DES data. Figure~\ref{fig:chi2-binw-dc345} shows the $\chi^2_\mathrm{CMB}$ distribution in the DC3, DC4 and DC5 chains, including DES-Y3 cosmic shear data, and using the EDE and EDE+$w(z)$ models.

\section{Principal Component Analysis of Late Dark Energy}
\label{sec:pca}
\begin{figure*}
    \centering
    \includegraphics[width=\textwidth]{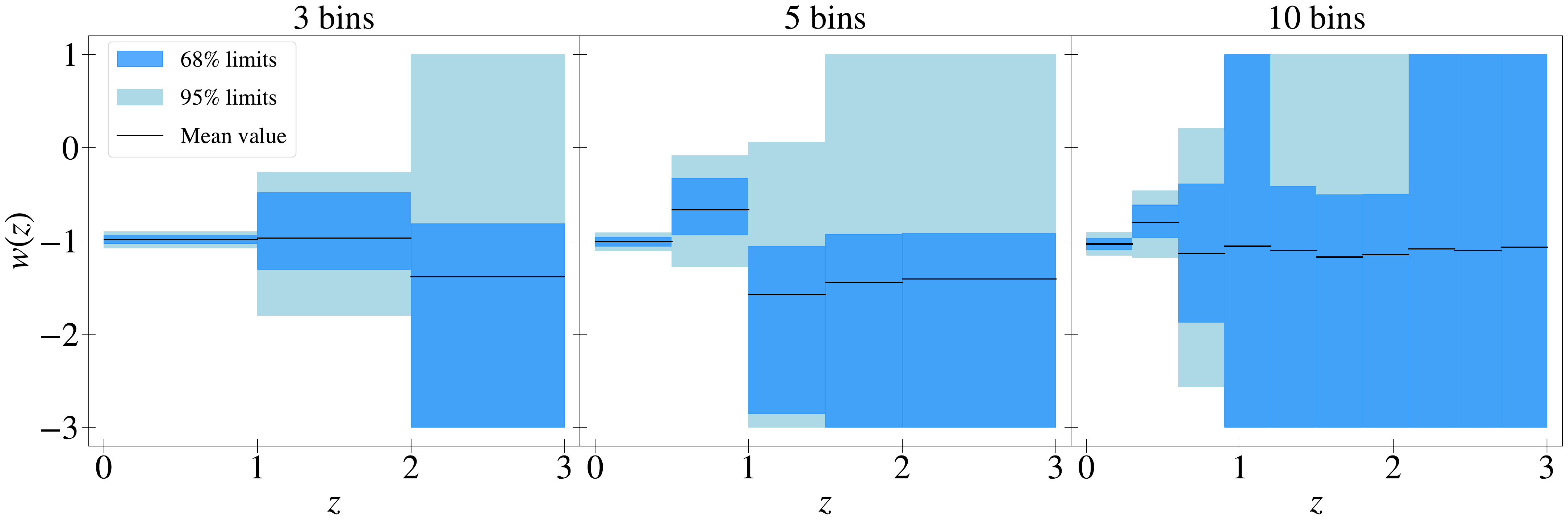}
    
    \caption{Mean (black line), 68\%  and 95\% limits (darker and lighter blue contours respectively) of the constraints for $w_i$, the dark energy equation of state within redshifts $z \in [z_i, z_{i+1} ]$. Vertical widths represent the confidence limits and the horizontal widths represent the chosen redshift binning.  The constraints are for the dataset combination DC1 (ACT TT, TE, EE spectra with $\ell_\mathrm{max} = 650$, Planck 2018 TT spectrum with $\ell_\mathrm{max} = 650$ + Planck 2018 low-$\ell$ EE spectrum, BAO from 6dFGS, SDSS DR7 main galaxy sample and SDSS BOSS DR12 consensus sample, Pantheon supernovae and Planck 2018 CMB Lensing spectrum) combined with DES-Y3 cosmic shear. We remind our MCMC priors $w_i \in [-3, 1]$ for $i > 0$. For redshifts $z > 1$, we have no constraining power.\textbf{Left panel}: constraints using the 3-bin model, with bin limits $z = [1,2,3]$. \textbf{Middle panel}: same constraints for the 5 bin analysis, where the bin limits are $[0.5, 1.0, 1.5, 2.0, 3.0]$. \textbf{Right panel}: same constraints for the binning $z_i = 0.3 \times i, i=1,...,10$.}
    \label{fig:reconstruction}
\end{figure*}

As a byproduct of our analysis, we present in this appendix constraints on the smooth, late-time dark energy behavior using our baseline dataset DC1, including DES-Y3 cosmic shear data. Figure~\ref{fig:reconstruction} shows constraints for $w(z_i)$, the dark energy equation of state at each redshift bin $[z_i, z_{i+1}]$ for the 5 and 10-bin cases. As expected, above $z = 1$ our constraining power decreases to the point that our constraints have no statistical significance. All constraints are well in agreement with a cosmological constant at late-times.

\begin{figure*}
    \centering
    \includegraphics[width=\textwidth]{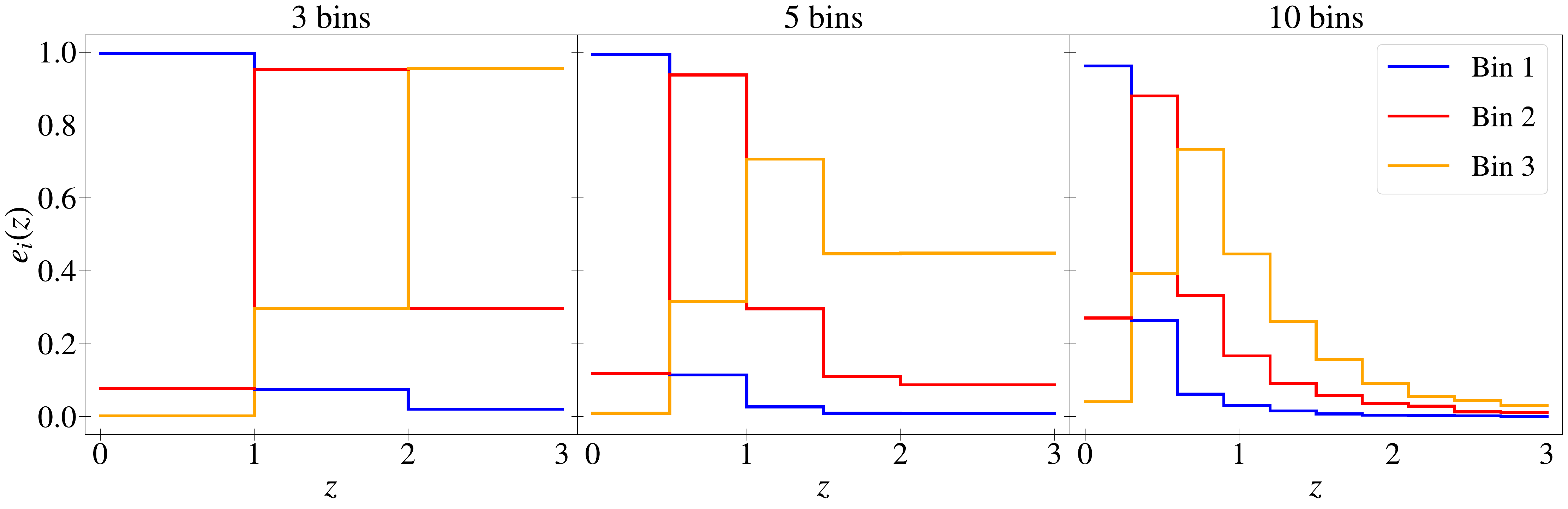}
    \caption{First (ordered by least variance) three $w(z)$ principal components, $e_i(z)$, $i = 1,2,3$. The PCs are constructed using the MCMC sample from DC1 (ACT TT, TE, EE spectra with $\ell_\mathrm{max} = 650$, Planck 2018 TT spectrum with $\ell_\mathrm{max} = 650$ + Planck 2018 low-$\ell$ EE spectrum, BAO from 6dFGS, SDSS DR7 main galaxy sample and SDSS BOSS DR12 consensus sample, Pantheon supernovae and Planck 2018 CMB Lensing spectrum) combined with DES-Y3 cosmic shear. The components of each PC are focused on each redshift bin, which we use to label the PCs. \textbf{Left panel}: PCs for the 3 bin analysis, where the bin limits are $[1, 2, 3]$ . \textbf{Middle panel}: First three PCs for the 5 bin analysis, where the bin limits are $[0.5, 1.0, 1.5, 2.0, 3.0]$.\textbf{Right panel}: First three PCs for the 10-bin analysis, where the bin limits are $z_i = 0.3 \times i, i=1,...,10$.}
    \label{fig:pcas}
\end{figure*}

Since the values of $w(z_i)$ are correlated, we perform a Principal Component Analysis to find uncorrelated parameters. The procedure is as follows: from the MCMC sample, we compute the sample covariance matrix for the equations of state in each bin, $w(z_i)$. This matrix is real and symmetric, and can therefore be diagonalized. The principal components are the eigenvectors $e_j$ of the sample covariance matrix. In the principal component basis, the covariance matrix is diagonal and thus the components are uncorrelated with each other. Thus, the dark energy equation of state can be expanded in the Principal Component basis as:
\begin{equation}
    w(z) = -1 + \sum_{i=1}^{N_\mathrm{bins}}\alpha_i e_i(z),
\end{equation}
where $N_\mathrm{bins}$ is the number of redshift bins and $\alpha_i$ is the component of $w(z)$ in the Principal Component $e_i$ direction. Figure~\ref{fig:pcas} shows the first three principal component basis functions. Each principal component is localized in a redshift bin but receive contributions from all other bins. The components $\alpha_i$ can then be interpreted as deviations from $w = -1$.

\begin{figure*}
    \centering
    \includegraphics[width=\textwidth]{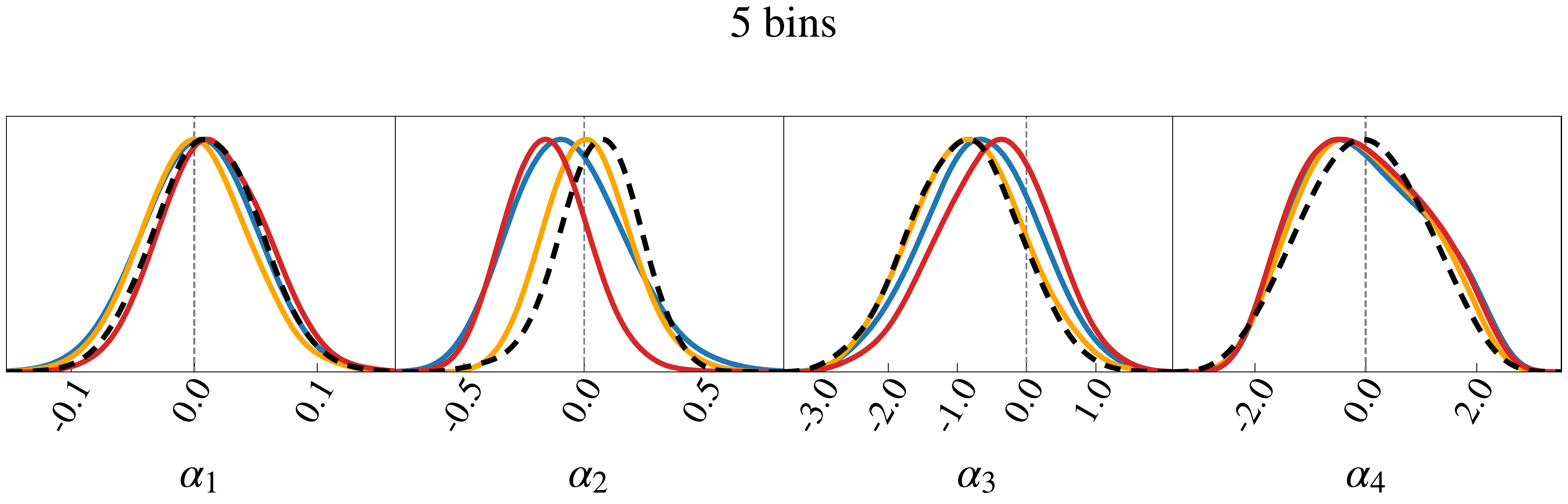}\\
    \includegraphics[width=\textwidth]{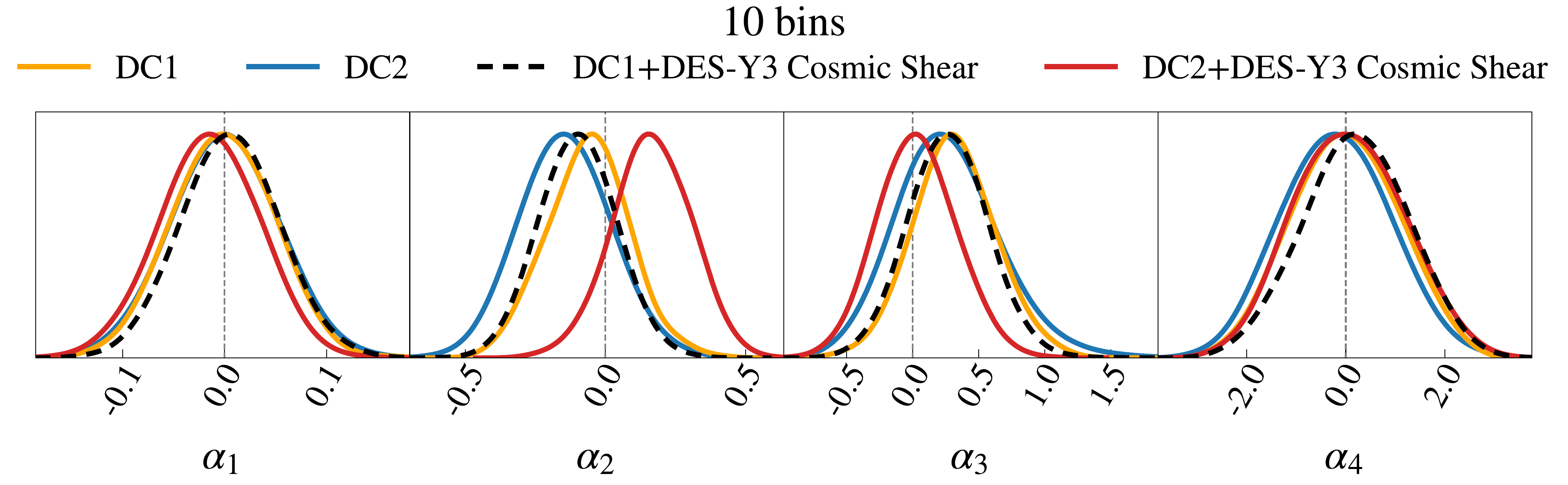}
    \caption{Marginalized posterior distributions of the first four principal component weights, $\alpha_i$, under the datasets DC1 (ACT TT, TE, EE spectra with $\ell_\mathrm{max} = 650$, Planck 2018 TT spectrum with $\ell_\mathrm{max} = 650$ + Planck 2018 low-$\ell$ EE spectrum, BAO from 6dFGS, SDSS DR7 main galaxy sample and SDSS BOSS DR12 consensus sample, Pantheon supernovae and Planck 2018 CMB Lensing spectrum) and DC2 (same as DC1, without CMB Lensing). We also show results including DES-Y3 cosmic shear data combined with DC1 and DC2. \textbf{Top Panel}: Constraints for the weights of the first four principal components for the 5 bin analysis. \textbf{Bottom Panel}: Same constraints as in top panel but for the 10 bin analysis.}
    \label{fig:pca_triangle}
\end{figure*}

Figure~\ref{fig:pca_triangle} shows marginalized constraints for the components $\alpha_i$ of the three first principal components. All constraints are compatible with the fiducial $\Lambda$CDM model.

\acknowledgments

JR acknowledges the financial support from FAPESP under grant \#2020/03756-2, São Paulo Research Foundation (FAPESP) through ICTP-SAIFR. The work of DHFdS is supported by a CAPES fellowship. RR is partially supported by a CNPq fellowship.
RR thanks INCT do e-Universo, LIneA and Fapesp for support. TE is supported by the Department of Energy grant DE-SC0020215. EK is supported by the Department of Energy grant DESC0020247. Simulations in this paper use High Performance Computing (HPC) resources supported by the University of Arizona TRIF, UITS, and RDI and maintained by the UA Research Technologies department. The authors would also like to thank Stony Brook Research Computing and Cyberinfrastructure, and the Institute for Advanced Computational Science at Stony Brook University for access to the high-performance SeaWulf computing system, which was made possible by a $\$1.4$M National Science Foundation grant ($\# 1531492$). The authors also thank the Laboratório Nacional de Computação Científica (LNCC) for our access and usage of the Santos Dumont high-performance computing system.

\bibliographystyle{JHEP}
\bibliography{early_late.bib}

\end{document}